\documentclass[aps,pra,showpacs,twoside,twocolumn,10pt,longbibliography,floatfix]{revtex4-2}
\usepackage[dvipsnames]{xcolor}
\usepackage[colorlinks=true, linkcolor=blue, citecolor=NavyBlue, urlcolor=blue]{hyperref}
\usepackage{url}  
\usepackage{wrapfig,graphicx}

\usepackage{epsfig,newlfont,amssymb,amsfonts,amsmath,bm,subfigure,palatino,mathtools,amsthm,braket,times,soul,enumitem,color}
\newcommand{\stkout}[1]{\ifmmode\text{\sout{\ensuremath{#1}}}\else\sout{#1}\fi}

\begin{document}
\raggedbottom

\title{Entanglement marker for lifetime of time crystal in transmon-modulated open Dicke model}

\author{Tanaya Ray$^1$, Shuva Mondal$^1$, and Ujjwal Sen$^1$ }

\affiliation{$^1$Harish-Chandra Research Institute,  A CI of Homi Bhabha National Institute, Chhatnag Road, Jhunsi, Prayagraj 211 019, India}

\begin{abstract} 
We investigate the discrete time crystal (DTC) phase in a qubit ensemble, periodically driven by its interaction with either a photon or a transmon field, which is prone to dissipative leakage. 
We find this DTC to be robust against changes in detuning and anharmonicity of the field mode. 
Additionally, we study the system in the semiclassical limit, where mean-field approximations are valid, and demonstrate the utility of a suitable semiclassical Hamiltonian for this purpose. 
Intriguingly, we observe that the system exhibits a transient DTC even with only two qubits. 
We  examine the dynamics of bipartite entanglement between the qubits and the field.
Our findings show that the entanglement saturates to a steady value early in the dynamics, following  a sudden peak. 
We find a strong positive correlation between this long-term entanglement value and the lifetime of the transient DTC,  
in a wide range of the parameter regime where the field is due to a lossy photon or a lossy transmon mode, with small detuning.
\end{abstract}

\maketitle

\section{Introduction}

\label{sec:intro}

The central question posed in the seminal paper of Wilczek~\cite{dtc-Wilczek-2012} is whether time-translation symmetry can be spontaneously broken, similar to the broken space-translation symmetry in crystals.
This exploration has led to the discovery of time crystals~\cite{dtc-Wilczek-2012, dtc-Shapere-2012, dtc-bruno-2013, dtc-Watanabe-2015, dtc-Keyserlingk-2016, dtc-khemani-2019, dtc-gong-2018, dtc-jiang-2011, dtc-ponte-2015, dtc-lazarides-2015, dtc-moessner-2017, dtc-sacha-2015, dtc-khemani-2016,dtc-else-2016,dtc-yao-2017,dtc-Russomanno-2017,dtc-ho-2017,dtc-huang-2018,dtc-matus-2019,dtc-kshetrimayum-2020,dtc-estarellas-2020,dtc-Maskara-2021,dtc-Wang-2021,dtc-Pizzi-2021,dtc-Collura-2022,dtc-Huang-2022,dtc-bull-2022,dtc-deng-2023,dtc-Liu-2023,dtc-huang-2023,dtc-Giergiel-2023}, a novel out-of-equilibrium state of matter acting as a `stable, conservative, macroscopic clock'~\cite{dtc-khemani-2019}. 
Symmetries have played a pivotal role in uncovering the novel features of this newfound state of matter. 
To be more precise, spontaneously-broken discrete time-translation symmetry is a hallmark of time crystals. 
Detailed investigations to converge upon the criteria to define a time crystal have been conducted, leading to one very important `no-go' theorem  ~\cite{dtc-Watanabe-2015, dtc-bruno-2013}, stating that 
the required non-trivial time-dependence featuring a time crystal can not be obtained in either the ground or the thermal state for a purely time-independent Hamiltonian. 
However, Floquet Hamiltonians, i.e., periodically-driven Hamiltonians, turn out to be ideal candidates for realising time crystals. 
Researchers have recently succeeded in experimentally realising such Floquet time crystals~\cite{dtc-zhang-2017,dtc-choi-2017,dtc-pal-2018,dtc-Rovny-2018,dtc-Smits-2018,dtc-Randall-2021,dtc-Kesler-2021,dtc-Xu-2021,dtc-Kyprianidis-2021,dtc-Taheri-2022,dtc-Mi-2021,dtc-frey-2022,dtc-bao-2024,dtc-shinjo-2024,dtc-liu-2024}. 
As revealed by recent experiments, superconducting systems hold great promise in realising time crystals and utilising them in various tasks such as crafting quantum clocks~\cite{dtc-Carraro_Haddad-2024, dtc-Surace-2019}, conducting quantum metrology~\cite{dtc-Lyu-2020,dtc-Montenegro-2023}, performing quantum simulations~\cite{dtc-estarellas-2020}, and designing quantum engines~\cite{dtc-carollo-2020}, among many others applications~\cite{dtc-Bomantara-2018, dtc-Engelhardt-2024, dtc-Iemini-2024}. 
The Floquet time crystals have been further categorised into a few varieties, the discrete time crystal (DTC) being possibly the earliest and arguably the most researched one.
Researchers have also realised the discrete time crystal phase in quantum processors built upon superconducting qubits~\cite{dtc-frey-2022,dtc-Xu-2021,dtc-Zhang-2022}.
A DTC features a discrete time-translation symmetry breaking, and infinitely-long-lived oscillations in the thermodynamic limit. 
Relaxing the condition of infinitely-long-lived oscillations entails the class of `prethermal' time crystals, where the late-time oscillations scale only exponentially even in the thermodynamic limit~\cite{dtc-Kyprianidis-2021,dtc-Natsheh-2021,dtc-Vu-2023}.  
Time quasi-crystals can be realised with quasiperiodic Hamiltonians, in contrast to Floquet driving required for the case of discrete time crystals~\cite{dtc-Flicker-2018,dtc-Dumitrescu-2018,dtc-Autti-2018,dtc-Huang2-2018, dtc-Giergiel-2018,dtc-Giergiel-2019,dtc-he-2024}.
A new member of the class of this out-of-equilibrium many-body phase of matter is the set of boundary time crystals, where continuous time-translation symmetry breaking occurs only in a macroscopic fraction of a many-body quantum system~\cite{dtc-Iemini-2018,dtc-Carollo-2024,dtc-Prazeres-2021,dtc-Piccitto-2021,dtc-Montenegro-2023,dtc-Cabot-2024,dtc-cabot-2023,dtc-gribben-2024}. 


The remaining part of the paper is structured as follows. In section~\ref{sec:dtc} of this work, we describe the key aspects describing a DTC.
Previous studies have established the presence of DTC in circuit QED systems, even in the presence of detuning~\cite{dtc-gong-2018}. 
The successful implementation of new-age quantum computers built upon superconducting elements and their versatile applications motivate further exploration into the quantum correlations at play in self-organising many-body quantum systems, leading to the emergence of discrete time crystals in such systems.
In this work, we study a chain of $N$ ($N$=2,3,4) 
qubits in a superconducting transmon field, connected to a lossy cavity, to establish the relation between entanglement and lifetime of a discrete time crystal. 
The Hamiltonian and relevant parameters are mentioned in section~\ref{sec:model} of this article. 
In section~\ref{sec:model-dtc}, we briefly discuss how periodic driving induces parity inversion in the described system, instigating the period-doubling behaviour. 
Temporal oscillations in this system can persist infinitely, as revealed through the semiclassical treatment of the system, described in section~\ref{sec:semiclassical}. But in practice, we will often need to resort to relatively small systems, and appeal to their possible quantum advantage and applicability in so-called `NISQ' (noisy intermediate-scale quantum) devices - superconducting systems 
being one of the most promising platforms for such devices. 
However, the advantage of new-era quantum devices made with nano-scale systems accompanies the drawback of a finite lifetime of a DTC phase realised in such systems.
In the succeeding section of this work, i.e. in section~\ref{sec:ent-lifetime}, we study the described system in the deep quantum regime of 2,3, and 4 qubits, and reveal the interplay between the  entanglement and lifetime of the DTC phase. 
We find that varying the system  size does not change the results qualitatively.
In this work, we therefore argue that entanglement can be a faithful marker to estimate the lifetime of DTC in finite-sized systems. 
We conclude the study in section~\ref{sec:conclusion}.

\begin{figure}
    \centering
    \includegraphics[width=0.3\textwidth]{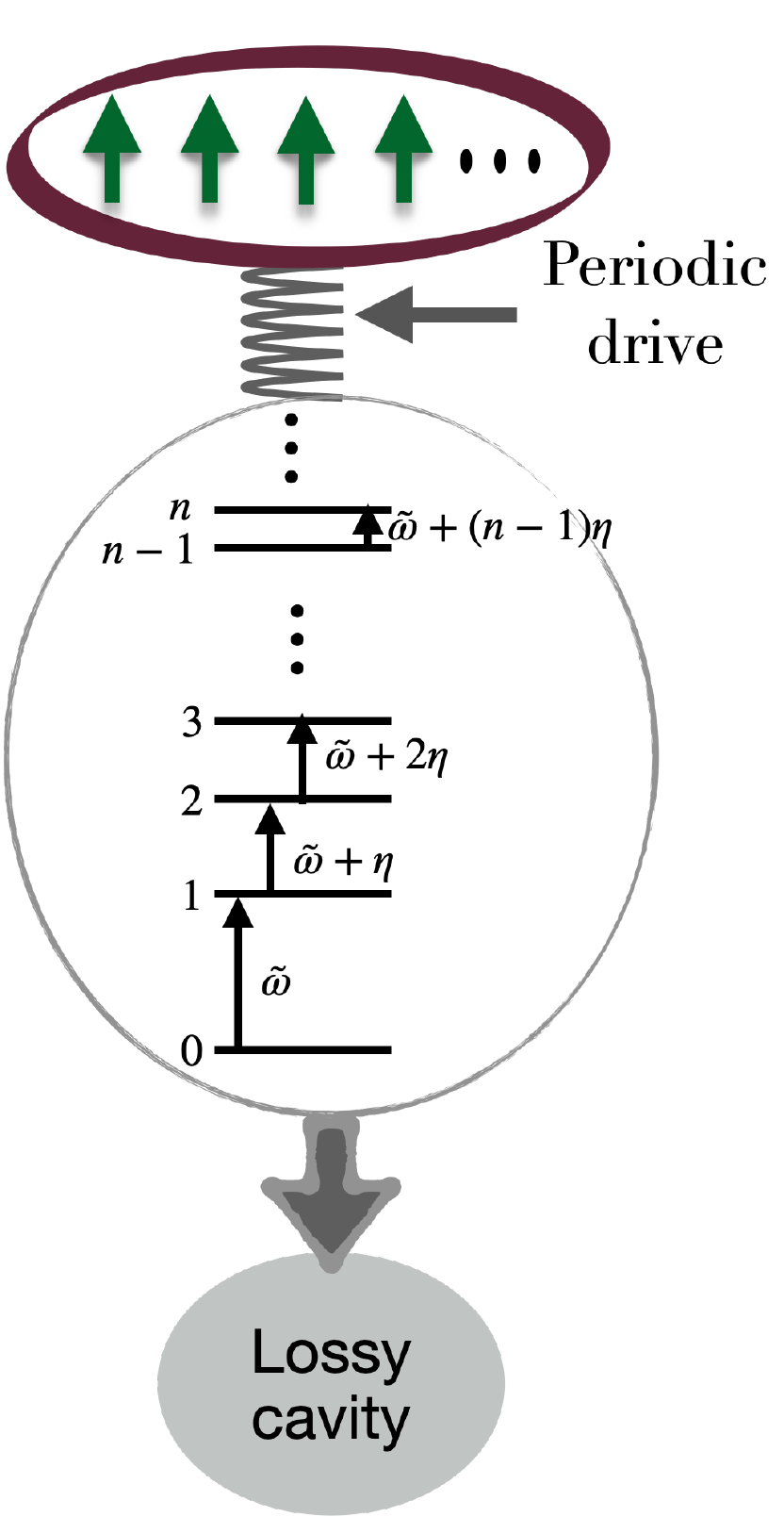}
    \caption{System under study: An ensemble of qubits, driven periodically via its interaction with a transmon field, prone to dissipative loss. Here, $\eta$ captures the anharmonicity in the field mode. For the case of a transmon, $\eta$ is negative, resulting in decreased energy spacings as one climbs up the energy ladder.}
    \label{fig:system}
\rule{\hsize}{.1pt}
\end{figure} 

\section{What defines a Discrete Time Crystal?}
\label{sec:dtc}

Discrete time crystal can emerge in a many-body Hamiltonian with discrete time-translation symmetry, under specific conditions, as will be discussed briefly in this section. 
In a DTC, as a result of broken discrete time-translation symmetry, the observables exhibit subharmonic response, i.e., oscillate with a time period that is multiple of the driving period in the Floquet Hamiltonian. 
But only a broken time-translation symmetry and infinitely long-lived oscillations are not enough to categorise such a out-of-equilibrium many-body state as a DTC. 
To qualify as a DTC, the system must also show crypto-equilibrium, and robustness, besides a broken discrete time-translation symmetry. 
We discuss these hallmarks of a DTC phase very briefly in the following.
\\
\textbf{Discrete time-translation symmetry breaking:}
Suppose the Floquet Hamiltonian $H(t)$ has a discrete time-translation symmetry with period of time $T$, so that $H(t+T)=H(t)$. 
In a system where this discrete time-translation symmetry is broken, we can find an operator $Q(t)$ evolving with a time period $T'$ such that $T'=nT$, where integer $n>1$. 
It is worth mentioning that such symmetry breaking is accompanied by rigid long-range spatio-temporal order. 
This in turn ensures that the non-trivial time-crystalline order does not result from the dynamics of individual particles but rather from the collective synchronisation of many strongly interacting degrees of freedom.
\\
\textbf{Crypto-equilibrium:} We say that a system is in crypto-equilibrium, when there exists some (possibly time-dependent) frame of reference from which the system is indistinguishable from a system in equilibrium, if measured stroboscopically, i.e., measured at times in integer multiples of the driving period. 
One signature of a DTC is that the system remains in a crypto-equilibrium, and as a result, no entropy is generated by the late-time oscillations. 
The criterion of crypto-equilibrium sets a DTC apart from a large class of other oscillating systems~\cite{dtc-Yao-2018}.
\\
\textbf{Robustness:} In a DTC, we can find an operator $Q(t)$ such that $Q(t+T')=Q(t)$ with $T'=nT$  and $n>1$, where $T$ is the driving period in the Floquet Hamiltonian. 
The demand of robustness in a DTC ensures that this $n$ should remain invariant over a measurable volume of initial conditions and local perturbations in system parameters, over a finite range of values.
\\
All these aforementioned criteria of a non-equilibrium many-body state of matter to qualify as a DTC can be equivalently summarised into the following definition. 
Discrete time crystal is a state of matter in which the Floquet eigenstates are necessarily entangled superpositions of macroscopically distinct states~\cite{dtc-Yao-2018}. 
With this we conclude our general remarks regarding the DTC, and we proceed to describe the specific system under study in the succeeding section. 

\begin{figure*}[t]
\begin{minipage}{0.33\textwidth}
 \centering 
     \includegraphics[width=\textwidth]{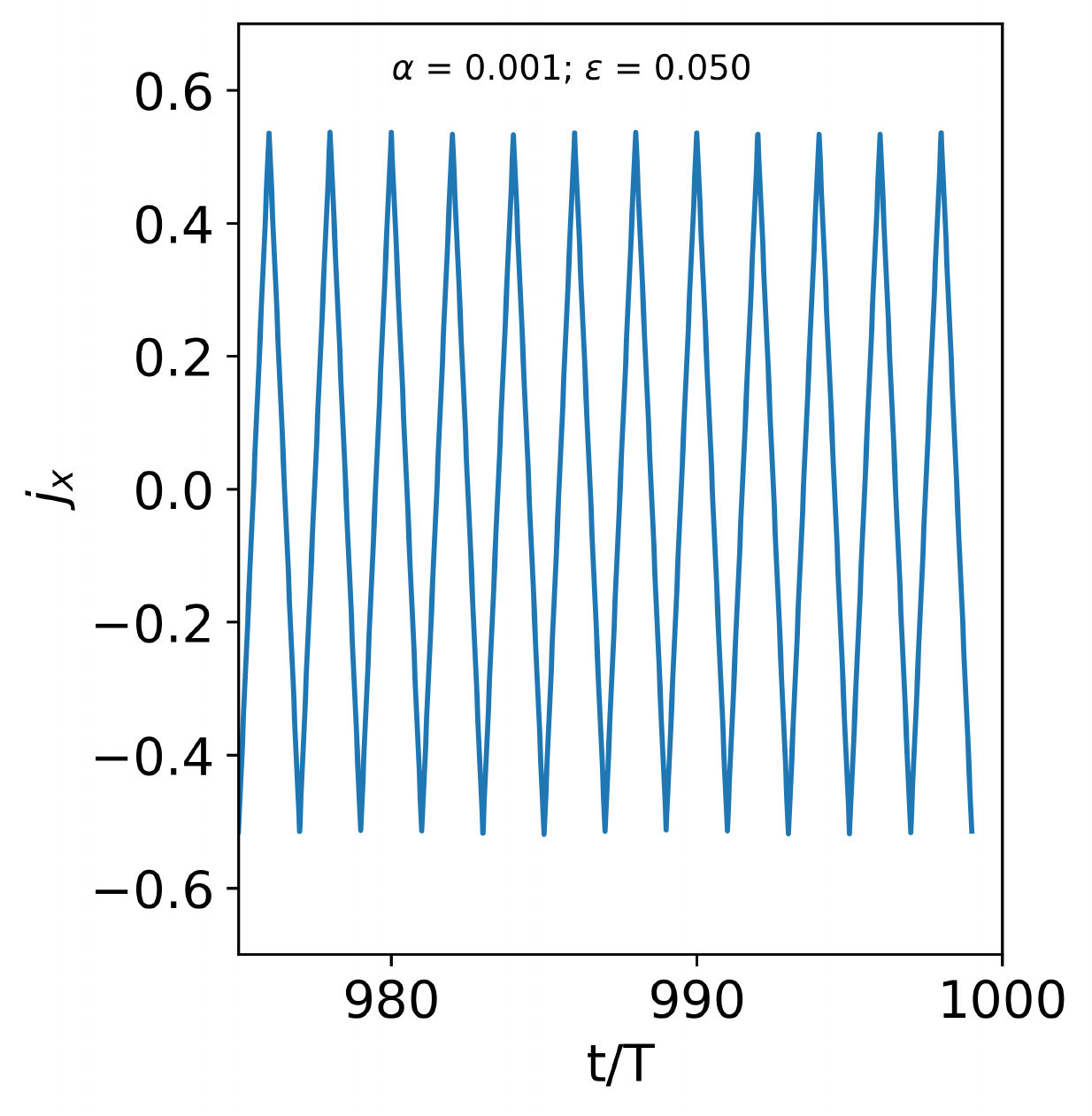}
    \centering
    \textbf{\phantom{xxxxx}(a)}
\end{minipage}
\hfill
\begin{minipage}{0.33\textwidth}
 \includegraphics[width=\textwidth]{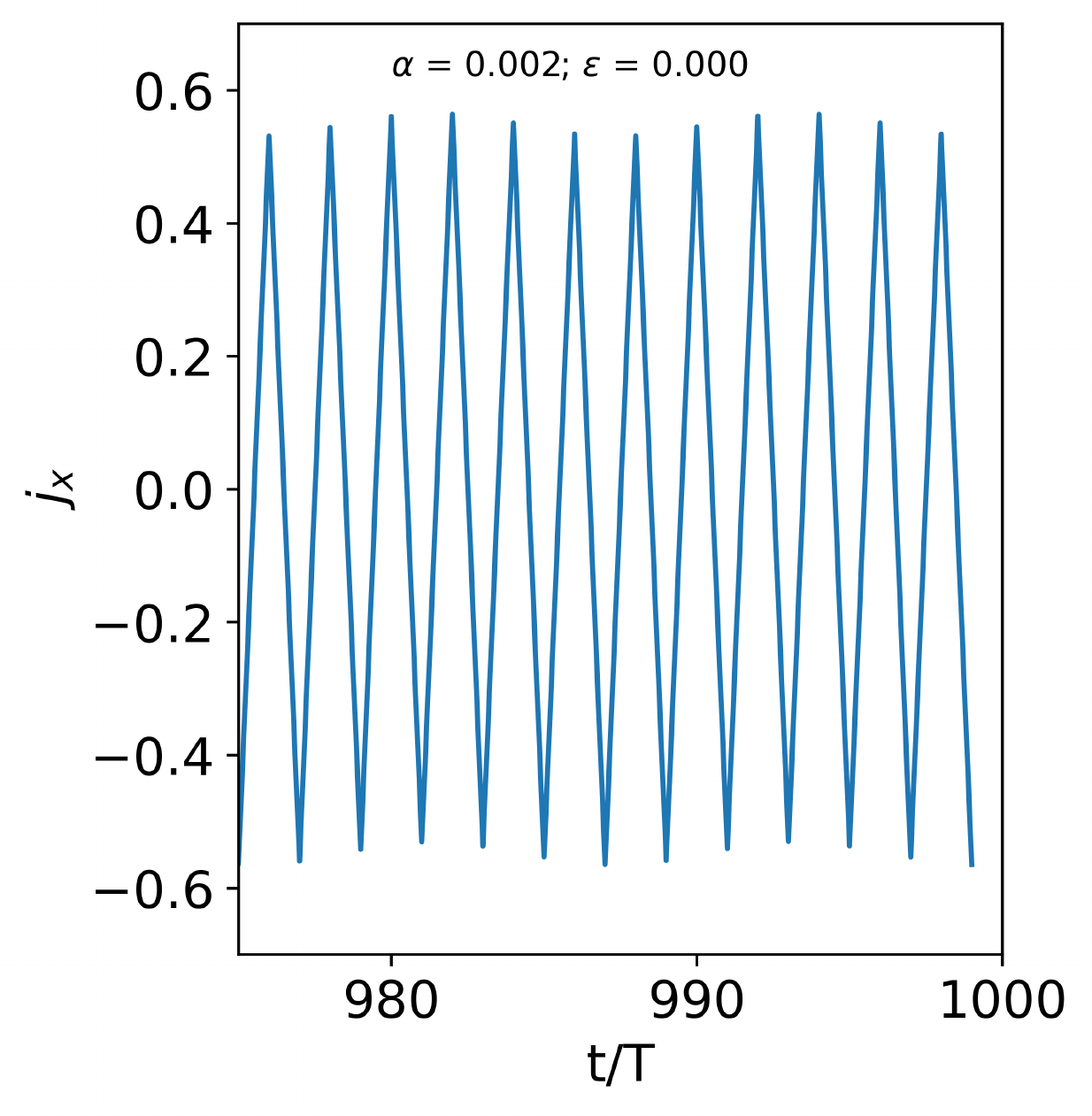}
    \centering
    \textbf{\phantom{xxxxx}(b)}
\end{minipage}
\begin{minipage}{0.33\textwidth}
 \includegraphics[width=\textwidth]{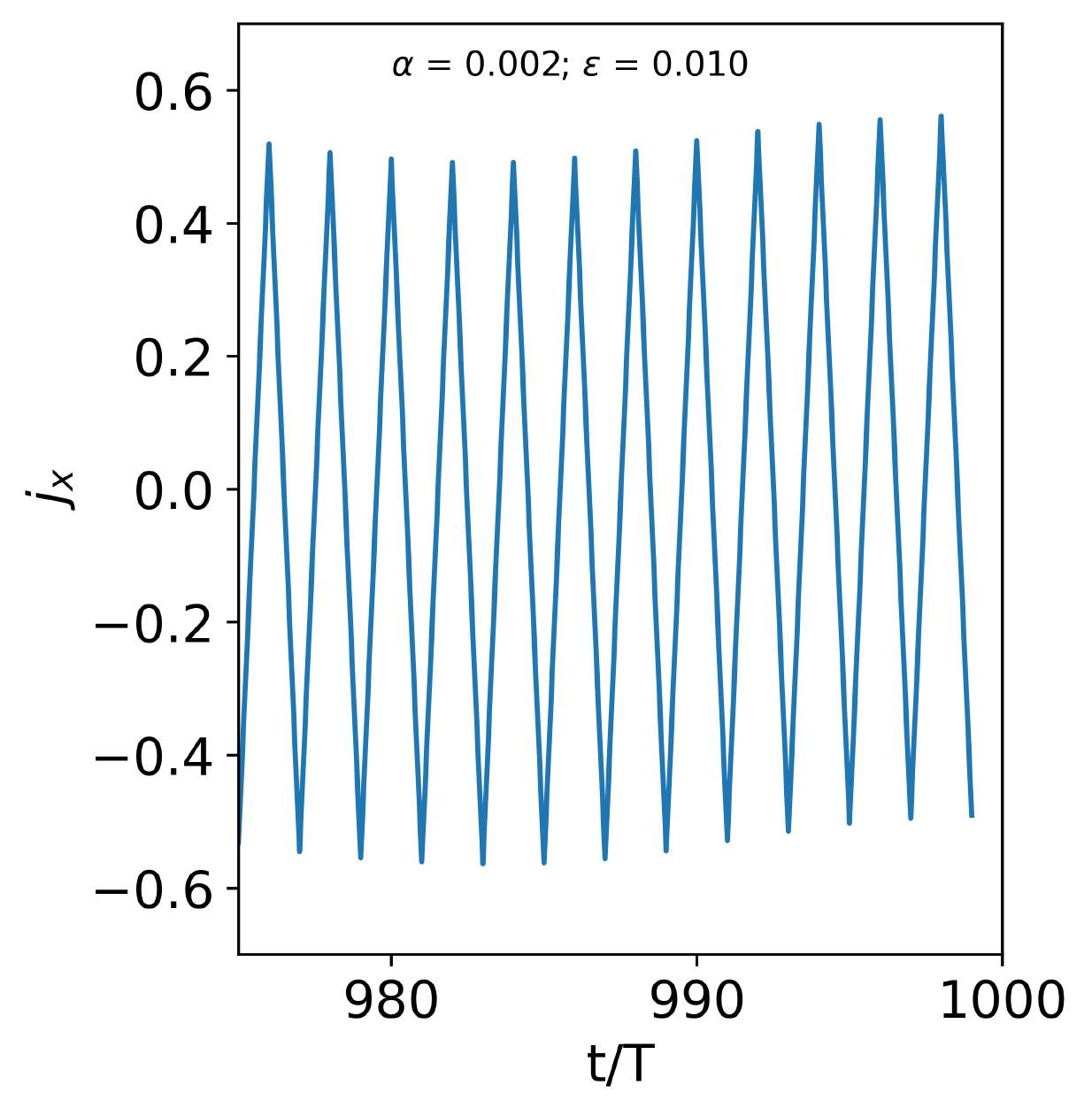}
    \centering
    \textbf{\phantom{xxxxx}(c)}
\end{minipage}
\caption{Semiclassical study revealing DTC in a system with Hamiltonian described by Eq.~\eqref{eq:semiclassical_hamiltonian}. The x-axis shows time in the units of the driving period $T$. Plotted along the y-axis are the stroboscopic dynamics of the average x-component of magnetisation, $j_x= \braket{J_x/N}$. Corresponding anharmonicity ($\alpha$) and detuning ($\epsilon$) are mentioned in the legends of the respective plots. Period-doubling behaviour of $j_x$, as revealed by the plots, is a signature of DTC. We have found this 2DTC to be robust against a range of $\alpha$ and $\epsilon$ and for a finite volume of initial conditions. The above plots are simulated starting from the initial condition as mentioned via Eq.~\eqref{eq:semiclassical-initial}.}
\label{fig:semiclassical_plots}
\rule{\hsize}{.1pt}
\end{figure*}

\section{Transmon-modulated open Dicke model}
\label{sec:model}

In this work, we scrutinise a qubit chain in transmon field, prone to dissipative leakage, for our study. 
Quite evidently, this model resembles the open Dicke model~\cite{dtc-dicke-1954,dtc-Kirton-2018,dtc-Rzaifmmode-1975,dtc-Emary-2003, dtc-hepp-1973,dtc-wang-1973,dtc-hioe-1973,dtc-Duncan-1974,dtc-Carmichael-1973}, but with a superconducting transmon field instead of a photonic field connecting the qubits to a leaky cavity.
The dynamics of such a system of $N$ qubits in a transmon field is governed by a master equation, expressed as~\cite{Petruccione, Alicki, Rivas, Lidar}:
\begin{equation}
    \label{eq:ME}
    \frac{d\rho_s(t)}{dt} = - i [H(\lambda,\eta, \epsilon),\rho_s(t)] + \kappa\mathcal{D}(a)\rho_s(t),
\end{equation}
where $a(a^{\dagger})$ is the annihilation(creation) operator of the field and the dissipator $\mathcal{D}(a)\rho_s(t) = a \rho_s(t) a^{\dagger} - \frac{1}{2} \{a^{\dagger} a, \rho_s(t)\} $ encapsulates the decay in presence of a Markovian bath with decay rate $\kappa$. 
Here, in Eq.~\eqref{eq:ME}, $\rho_s(t)$ features the density matrix of the system comprised of the qubits in the transmon field, and $H(\lambda,\eta, \epsilon)$ delineating the Hamiltonian of the same :
\begin{equation}
    \label{eq:H_s}
    H(\lambda,\eta, \epsilon) = H_f(\omega, \eta) + \omega_0 J_z + \frac{2 \lambda}{\sqrt{N}}(a+a^{\dagger})J_x
    ,
\end{equation}
the interaction between the qubit chain and superconducting field parameterized by the radiation-matter coupling 
 $\lambda$. 
 In the above equation, Eq.~\eqref{eq:H_s}, $ J_{\mu} = \frac{1}{2} \sum\limits_{j=1}^N \sigma_i^{\mu} (\mu = x, y, z)$ denotes the collective pseudospin operator,  $\omega_0$ is the atomic frequency signifying the transition energy of the qubits in the system, $\omega$ and $\eta$ together characterizes the field Hamiltonian $ H_f(\omega, \eta) $ described as following, and $\epsilon$ quantifies the detuning between the field and qubits via the couple of equations
$
    \omega = (1-\epsilon)\omega_T  \hspace{.7em}\& \hspace{.7em} \omega_0 = (1+\epsilon)\omega_T
    ,
$
where $\omega_T = \frac{2\pi}{T}$ is the frequency of the Floquet drive in Eq.~\eqref{eq:H_s}, provoking the qubit chain to interact with the field mode.

Now let us turn our attention to the field Hamiltonian $ H_f(\omega, \eta) $ and explore the limits on the superconducting field modes considered henceforth.
We have focused our work on the cases when the field mode is weakly anharmonic, attributing such anharmonicity to transmons~\cite{ctn_jj_17, ctn_jj_12, ctn_jj_13, ctn_jj_14, Ithier2005, Cottet2002, Houck2009}. 
As a result, we can consider the field Hamiltonian to be embodied in the following form: 
\begin{equation}
    \label{eq:h-tmon}
    H_f(\omega, \eta) =\omega \hat{a}^{\dagger} \hat{a}  -\dfrac{E_{C}}{12}        (\hat{a}_i^{\dagger}+ \hat{a}_i)^4 ,
\end{equation}
where $a^{\dagger}$($a$) is the field creation(annihilation) operator, and $\omega = \sqrt{8E_{C}E_{J}}$, with $E_C$ and $E_J$ being the charging energy and Josephson energy, respectively. $x$ in Eq.~\eqref{eq:h-tmon} is a function of the charging energy $E_C$, as mentioned in the following. 
Expanding the quartic term in Eq.~\eqref{eq:h-tmon}, and dropping the fast-rotating terms, we can reframe the transmon Hamiltonian in Eq.~\eqref{eq:h-tmon} as
\begin{equation}
    \label{eq:H-f}
    H_f(\omega, \eta) = (\omega +\frac{\eta}{2}) a^{\dagger} a + \frac{\eta}{2} (a^{\dagger} a)^2,
\end{equation}
where $\eta=-E_C$ captures the anharmonicity in the field.
Such a negative value of $\eta$ features a field mode of level-spacing decreasing with increasing energy levels, ascribing it to the weak anharmonicity in transmons. 
To encompass the behavior of the considered system in a transmon field, we need to confine our study in the parameter space where the anharmonicity ($\eta$) is roughly a couple of orders of magnitude smaller than the first excitation energy. 
This anharmonicity $\eta$ clearly quantifies the difference between any two consecutive transition energies, as one goes up in the energy ladder. 
The ground state to first excited state transition energy of this field mode is $\tilde{\omega}=\omega+\eta$, as expressed through the Hamiltonian $H_f$ of Eq.~\eqref{eq:H-f}. 
In the absence of detuning ($\epsilon=0$), the qubit energies and the first excitation energy of the transmon field becomes the same, i.e., $\tilde{\omega}=\omega_0$. 
And $\tilde{\omega} = \omega$ in the case of a photonic field, and hence consistent with the expression of detuning mentioned earlier in this section.
The energy levels in such a transmon field turn out to be 
\begin{equation} 
\label{eq:E-f}
E_n(\tilde{\omega}, \eta) = n\tilde{\omega} + \dfrac{n(n-1)}{2} \eta , \quad \forall n = 1, 2, 3, \cdots .
\end{equation} 
Evidently, we will get back the Dicke Hamiltonian with a photonic field in the limit $\eta\rightarrow 0$.
It should be noted that we have taken  $\hbar$ to be unity in all the preceding and succeeding discussions throughout this work.

\begin{figure*}[t]
\begin{minipage}{\linewidth}
\includegraphics[width=0.49\textwidth]{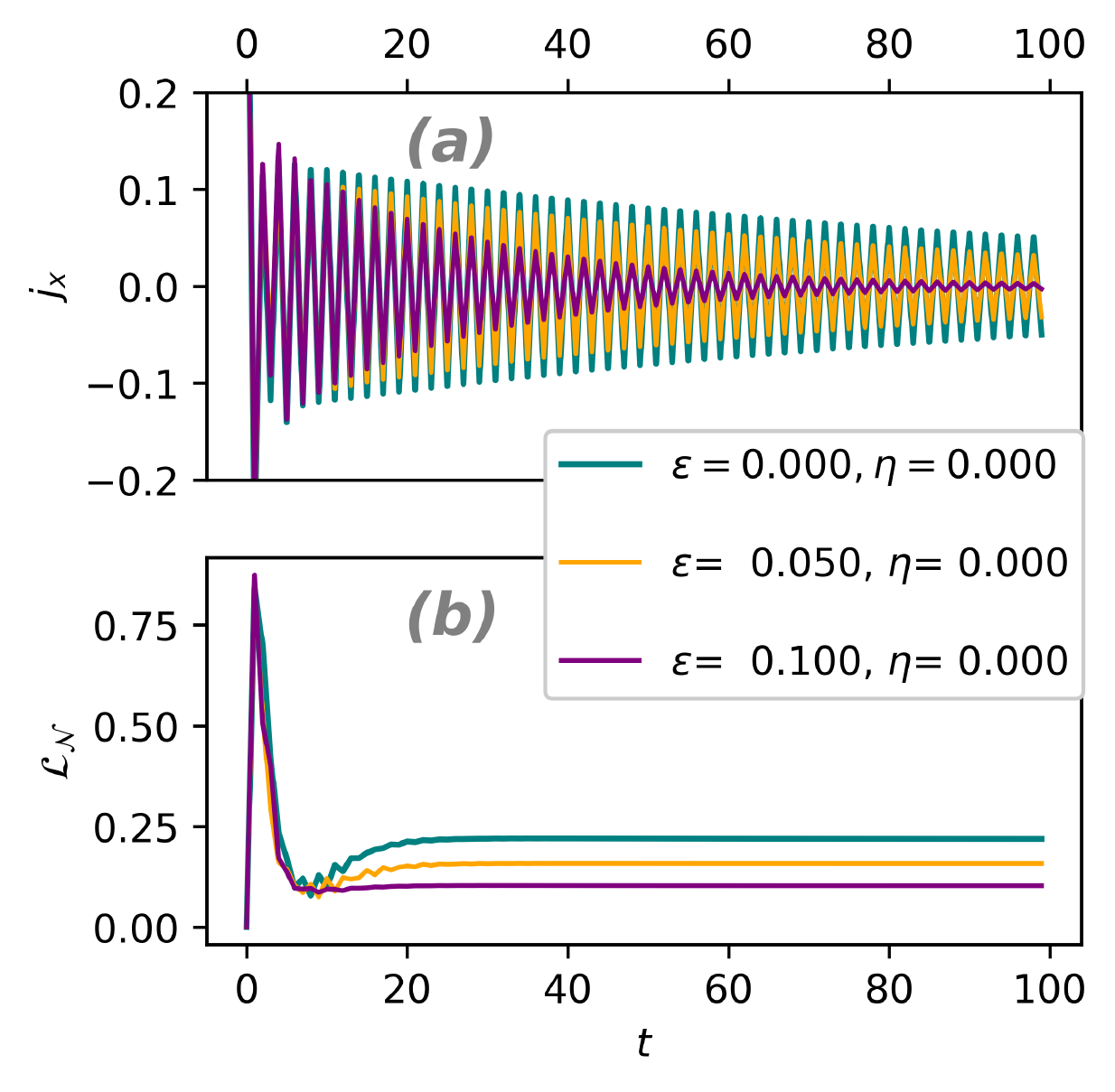}
\includegraphics[width=0.49\textwidth]{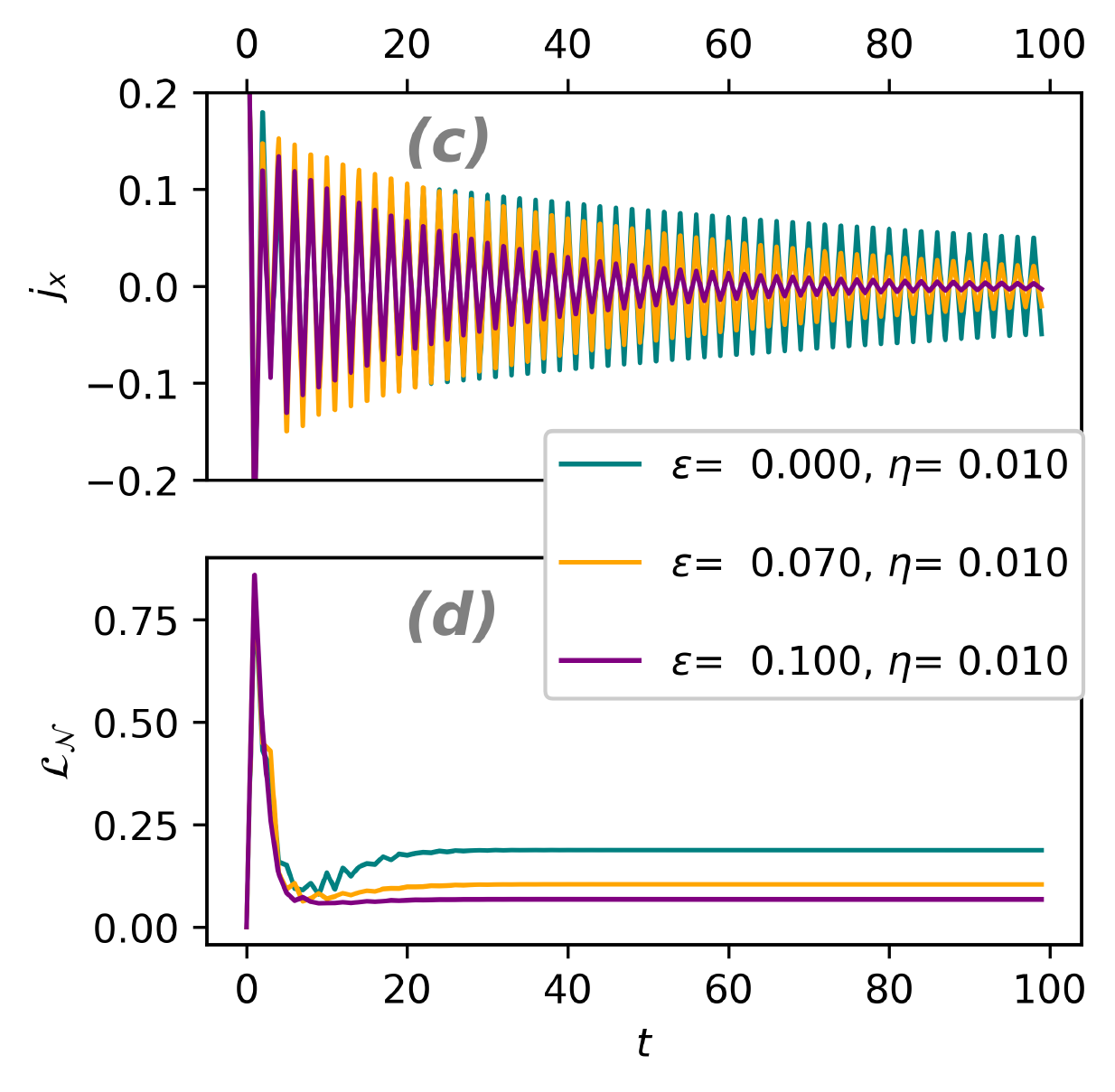}
\end{minipage}
\caption{ Stroboscopic dynamics of average magnetisation $j_x$ and entanglement, in a system of two qubits driven by a photon/transmon field mode, which in turn is prone to dissipative leakage. The field mode is truncated with 16 levels. Eq.~\eqref{eq:H_s} describes the Hamiltonian of the system under study.
The x-axis shows time in the units of the driving period $T$, in all the panels. In panels (a) and (c), plotted along the y-axis is the stroboscopic dynamics of the average x-component of magnetisation, $j_x= \braket{J_x/N}$ of the two-qubit system. In panels (b) and (d), logarithmic negativity, a quantifier of the bipartite entanglement between the two qubits and the field mode, is plotted along the y-axis, in the unit of e-bits. Corresponding anharmonicity ($\alpha$) and detuning ($\epsilon$) are mentioned in the legends of the respective plots. Period-doubling behaviour of $j_x$, as revealed by the plots, is a signature of DTC. We have found this DTC with period-doubling dynamics (2DTC) to be robust against a range of $\alpha$ and $\epsilon$ and for a finite volume of initial conditions. The above plots are simulated starting from the initial state $\ket{\Rightarrow}\otimes \ket{0}$, where $\ket{0}$ is the photon/transmon vacuum and $\ket{\Rightarrow}$ is the eigenstate of $J_x$ operator with eigenvalue $N/2$.
It should be noted here that $\eta$, as expressed through the Eq.~\eqref{eq:E-f}, is negative in the transmon regime and zero for photonic field mode. However, in the legends of the plots, we mention the absolute values of the anharmonicity parameter $\eta$. $\kappa$ =0.05 entails the effect of the dissipative leakage on the photon/transmon field.  }
\label{fig:dtc-e}
\rule{\hsize}{.1pt}
\end{figure*}

\begin{figure}[b]
\begin{minipage}{0.482\textwidth}
\centering
\includegraphics[width=\textwidth]{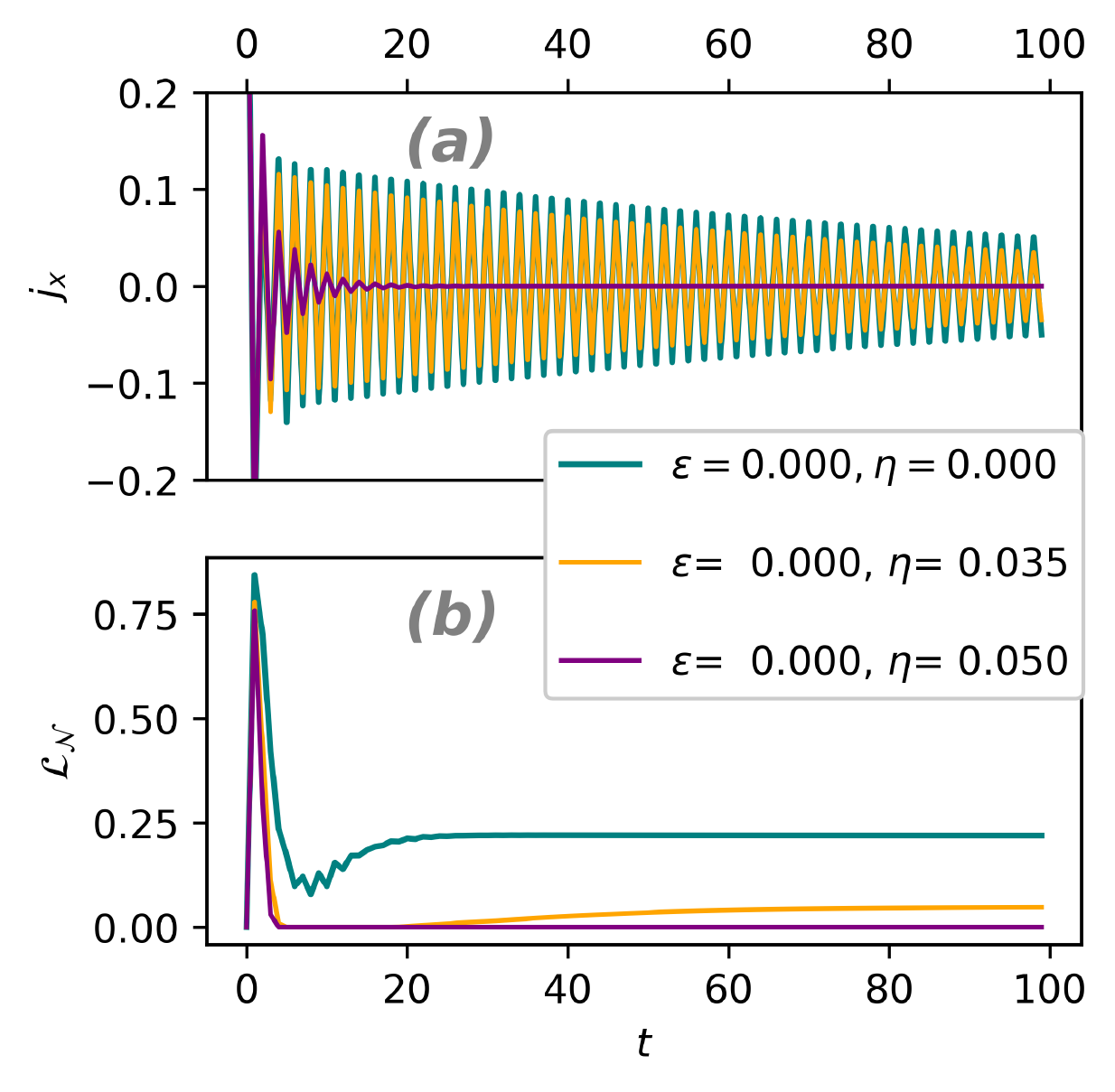}
\caption{Stroboscopic dynamics of average magnetisation $j_x$ and entanglement, in a system of two qubits driven by a photon/transmon field mode, prone to dissipative leakage. 
The x-axis shows time in the units of the driving period $T$, in all the panels. In panel (a), plotted along the y-axis is the stroboscopic dynamics of the average x-component of magnetisation, $j_x= \braket{J_x/N}$ of the two-qubit system. In panel (b), logarithmic negativity, a quantifier of the bipartite entanglement between the two qubits and the field mode, is plotted along the y-axis, in the unit of e-bits. All other considerations are same in mentioned in the caption of Fig.~\ref{fig:dtc-e}.
It should be noted here that $\eta$, as expressed through the Eq.~\eqref{eq:E-f}, is negative in the transmon regime and zero for photonic field mode. However, in the legends of the plots, we mention the absolute value of the anharmonicity parameter $\eta$.}
\label{fig:dtc-x}
\end{minipage}
\rule{\hsize}{.1pt}
\end{figure}

\begin{figure*}[t]
\begin{minipage}{\linewidth}
\includegraphics[width=0.49\textwidth]{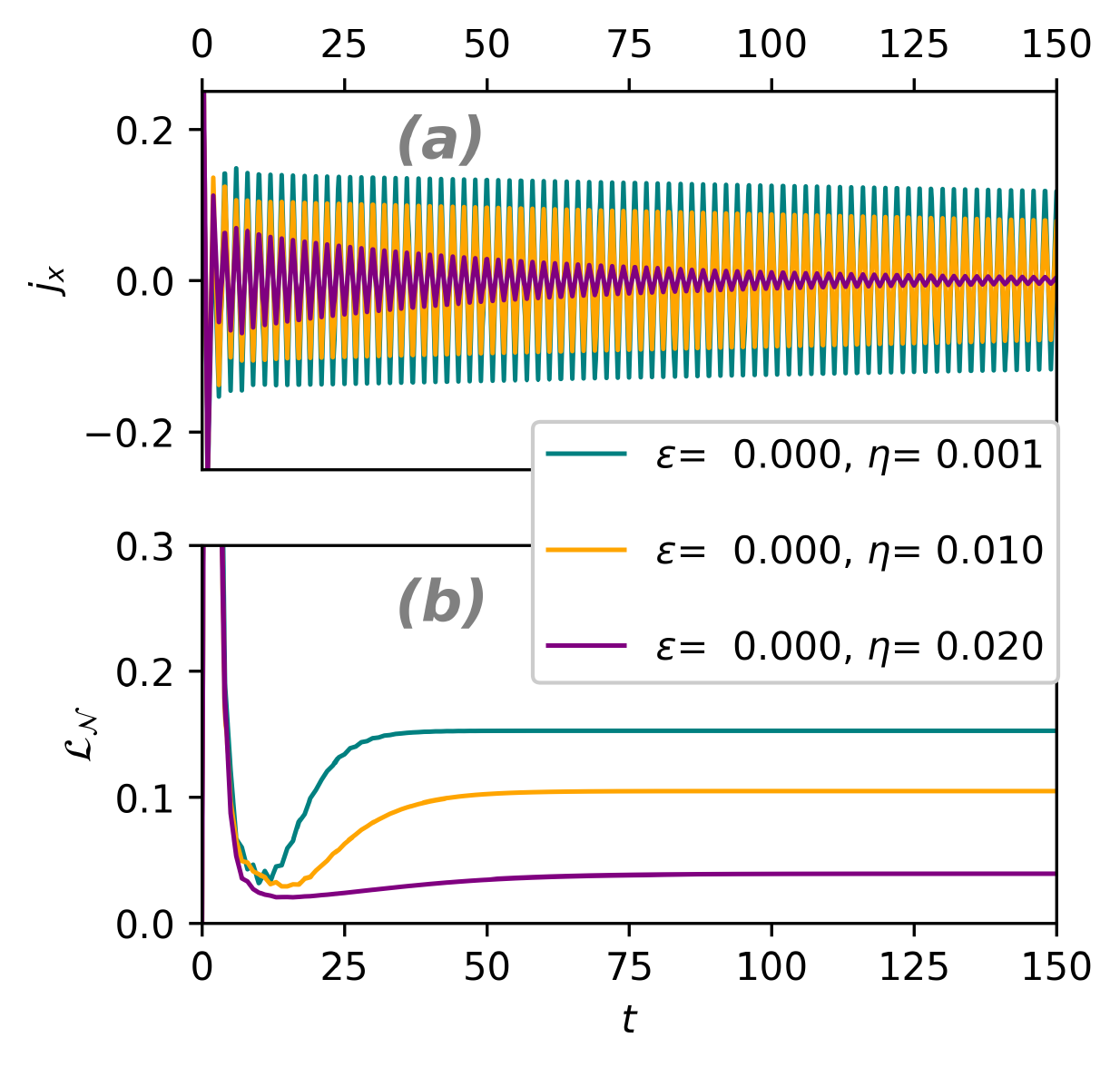}
\includegraphics[width=0.49\textwidth]{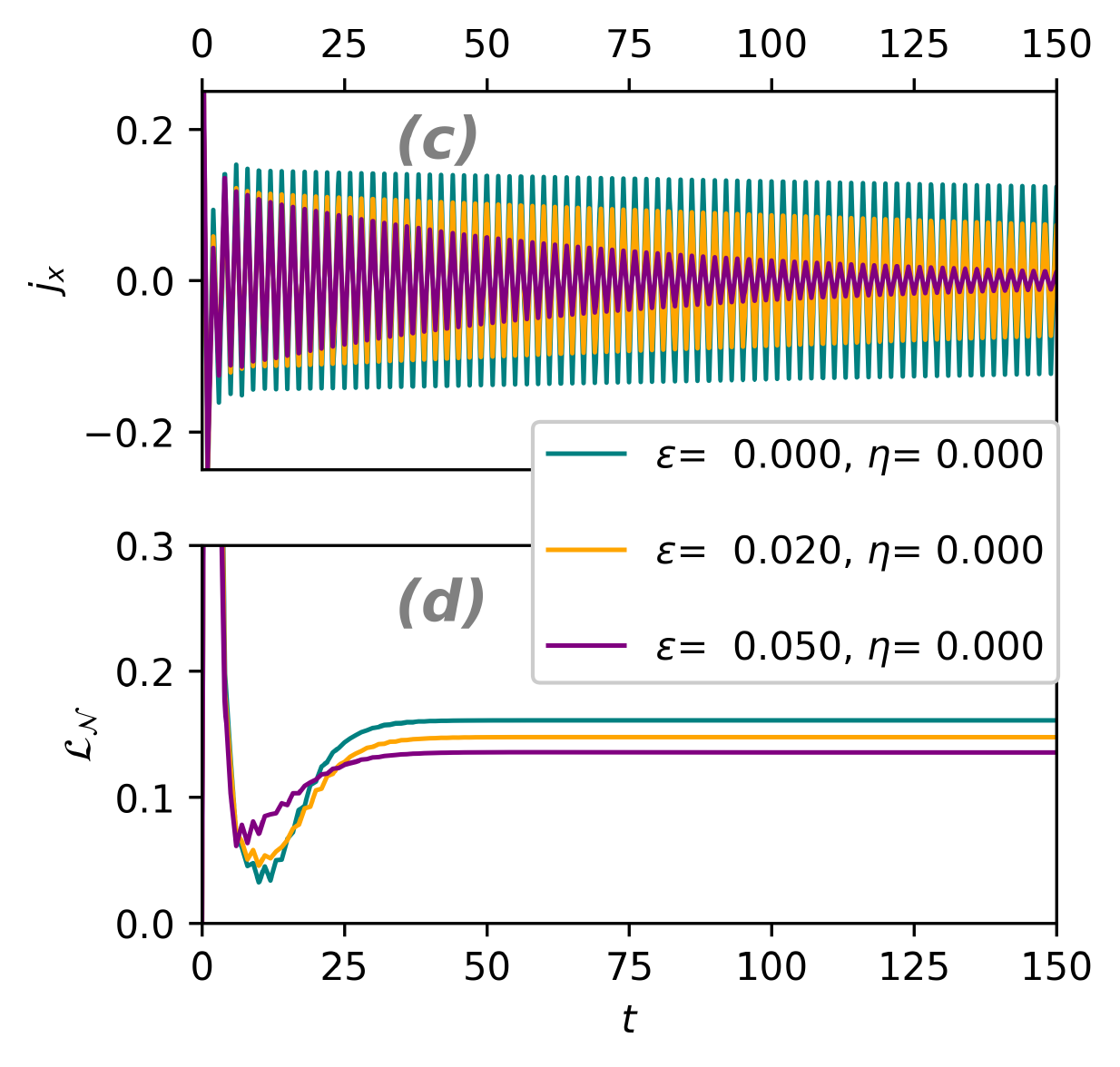}
\end{minipage}
\caption{ Stroboscopic dynamics of average magnetisation $j_x$ and entanglement, in a system of four qubits driven by a photon/transmon field mode, prone to dissipative leakage. The field mode is truncated at 16 levels. Eq.~\eqref{eq:H_s} describes the Hamiltonian of the system under study, with $N=4$.
The x-axis shows time in the units of the driving period $T$, in all the panels. In panels (a) and (c), plotted along the y-axis is the stroboscopic dynamics of the average x-component of magnetisation, $j_x= \braket{J_x/N}$ of the two-qubit system. In panels (b) and (d), logarithmic negativity, a quantifier of the bipartite entanglement between the four-qubit ensemble and the field mode, is plotted along the y-axis, in the unit of e-bits. Corresponding anharmonicity ($\alpha$) and detuning ($\epsilon$) are mentioned in the legends of the respective plots. Period-doubling behaviour of $j_x$, as revealed by the plots, is a signature of DTC. We have found this 2DTC to be robust against a range of $\alpha$ and $\epsilon$ and for a finite volume of initial conditions. The above plots are simulated starting from the initial state $\ket{\Rightarrow}\otimes \ket{0}$, where $\ket{0}$ is the field vacuum and $\ket{\Rightarrow}$ is the eigenstate of $J_x$ operator with eigenvalue $N/2$.
It should be noted here that $\eta$, as expressed through the Eq.~\eqref{eq:E-f}, is negative in the transmon regime and zero for photonic field mode. However, in the legends of the plots, we only mention the absolute values of the anharmonicity parameter $\eta$. $\kappa$ =0.05 entails the effect of the dissipative leakage on the photon/transmon field.  }
\label{fig:dtc-4}
\rule{\hsize}{.1pt}
\end{figure*}

\section{Emergence of DTC through Floquet driving}
\label{sec:model-dtc}

In this section, we aim to examine the time evolution of the system under study as described through Eq.~\eqref{eq:H_s}. In the absence of any anharmonicity in the field term, the system can be described by the Dicke Hamiltonian~\cite{dtc-dicke-1954,dtc-Kirton-2018,dtc-Rzaifmmode-1975,dtc-Emary-2003, dtc-hepp-1973,dtc-wang-1973,dtc-hioe-1973,dtc-Duncan-1974,dtc-Carmichael-1973,dtc-Dimer-2007}. : 
\begin{equation}
\label{Dicke Hamiltonian}
    H_{Dicke}=\omega a^{\dag} a+ \omega _{0} J_{z}+\frac{2 \lambda}{\sqrt{N}}(a^{\dag}+a)J_{x}.
\end{equation} 
Dicke model is an atomic ensemble in a high-finesse cavity~\cite{dtc-Duan-2000,dtc-Kurk-2021,dtc-Hammerer-2010}.
This model shows a DTC phase, accompanied by a $\mathbb {Z}_2$ symmetry breaking characterized by the parity operator~\cite{dtc-Emary-2003,dtc-Baumann-2010,dtc-Baumann-2011}. 
The Parity operator in Dicke Hamiltonian can be expressed as
\begin{equation}
\label{eq: parity operator}
    P \sim exp[-i\pi (a^{\dag} a+ J_{z})].
\end{equation}
Note that $P$ has eigenvectors $\ket{p;m,j}$ with eigenvalues  $exp[i\pi (p+ m)] = \pm 1.$ ($p$ denotes photon number, $m$ indicates eigenvalue of $J_{z}$ and $j$ indicates total angular momentum). Also, $P^{2}=\pm 1$. Thereby establishing $P$ to an appropriate parity operator in the mentioned context. 
It can also be shown that $[H_{Dicke},P]=0$, implying a $Z_{2}$ symmetry. 
It is well-established through previous studies that this $Z_{2}$ symmetry in the Dicke model is broken spontaneously beyond a critical driving strength $\lambda$: $\lambda >\lambda_{c}= \frac{1}{2}\sqrt{\omega\omega _{0}}$, referred to as the superradiant phase transition~\cite{dtc-Kirton-2018}, as described in the thermodynamic limit. 
This critical interaction strength is modified when we consider the open Dicke model, i.e., the Dicke model connected to a leaky cavity, described similarly as through the dissipator in the master equation Eq~.\eqref{eq:ME}. This new critical point is at 
$ \lambda_{c}=\frac{1}{2} \sqrt{\frac{\omega_{0}(\omega^{2}+\kappa^{2}/4)}{\omega}} $.
 Due to this breaking of $\mathbb {Z}_2$ symmetry we get a doubly degenerate ground state in the superradiant phase of the open Dicke model featured by $\lambda>\lambda_c$, in the limit $N \rightarrow \infty$.
Now, consider the time evolution operator of the Dicke model, with zero detuning and driving frequency : ($\omega = \omega_{0}=\frac{2\pi}{T}$). The time evolution over a half-time period is expressed as
$$U(T/2)=exp[-i\pi(a^{\dag} a+ J_{z})-\frac{i T \lambda}{\sqrt{N}}(a^{\dag}+a)J_{x}], $$
which turns out to be the aforementioned parity operator when $\lambda=0$.
Next, starting from very close to one of the degenerate ground states, the system is evolved under a periodic drive such that  $\lambda=\lambda_{0}$ for $nT<t\leq (n+\frac{1}{2})T$ and $\lambda=0$ for $(n+\frac{1}{2})T<t \leq (n+1)T$ for any integer $n$ and $\lambda_0>\lambda_c$. 
Clearly, the system's time evolution is continuously switched between two kinds of dynamics, depicted as follows. 
In the 1st half-period of time $T$, the system evolves in the vicinity of one of the two fixed points, i.e., the symmetry-broken states as mentioned before. 
In the second half of the period the time evolution acts as the parity operation as described apriori, so that the state of the system evolves to the other fixed point. 
As a result, the system recurrently visits the same fixed point with a period of $2T$. 
A balance between the driving strength and dissipation strength present in the open Dicke model plays a crucial role in sustaining this oscillating dynamics, delaying the thermalisation of the system substantially. 
Since the $x$ component of magnetisation distinguishes the mentioned symmetry-broken states, the time evolution of $\braket{J_x}$ reveals time-translation symmetry breaking in this system and the emergence of a discrete time crystal.

Careful investigation regarding the long-term behaviour of the considered system and robustness against the detuning $\epsilon$ has already established the time crystalline phase of this system, backed up by the dynamics revealed through the semiclassical treatment of the open Dicke model under periodic driving~\cite{dtc-gong-2018}. 
In this work, we also check the robustness of this 2DTC phase against anharmonicity in the field. 
Our study not only establishes the stability of this DTC phase against anharmonicity but also reveals an interesting relation between the entanglement and the lifetime of such a DTC phase, as will be discussed in section~\ref{sec:ent-lifetime}.

\section{Emergence of DTC in semiclassical limit}
\label{sec:semiclassical}

In this section, we study the considered system in the thermodynamic limit $N\rightarrow\infty$, where the relative fluctuations in the local observables become negligible, so that the semiclassical approach is justified~\cite{dtc-Bastidas-2012,dtc-Bhaseen-2012,dtc-chitra-2015}.
Also, as mentioned in the previous section, our operation regime is such that the system is in the superradiant phase, where $n(a^{\dagger}a)$ is proportional to the number of qubits $N$~\cite{dtc-wang-1973, dtc-Kirton-2018}. 
The transmon field is only weakly anharmonic, so that we expect the system to exhibit superradiance in the presence of this anharmonicity as well. 
To obtain meaningful quantities in the thermodynamic limit, we need to divide the anharmonicity inducing quartic term by $N$. 
We further justify this step in view of extensivity of the considered Hamiltonian, discussed at the end of this section.
Consequently, we reframe the Hamiltonian considered in Eq.~\eqref{eq:H_s} as
\begin{equation}
\label{eq:semiclassical_hamiltonian}
    H=\omega a^{\dag} a - \frac{\alpha}{N} (a^{\dag}+a)^{4} + \omega _{0} J_{z}+\frac{2 \lambda}{\sqrt{N}}(a^{\dag}+a)J_{x},
\end{equation}
where, \(\eta = - 12 \alpha/N\) captures the anharmonicity in the field mode.
In large $N$ limit, we can ignore fluctuation of product of operators around the product of their mean values. 
In terms of the average magnetisation $j_{\mu}={\braket{J_{\mu}}}/{N}$, and the scaled variables $x={\braket{(a+a^{\dag})}}/{\sqrt{2N\omega}}$, and $p=i{\braket{(a^{\dag}-a)}}/{\sqrt{2N/\omega}}$, the governing semiclassical equations become 
\begin{equation*}\label{eq: nJx2}
    \frac{d j_{x}}{dt}=-\omega_{0} j_{y},
\end{equation*}
\begin{equation*}\label{eq: nJy2}
    \frac{d j_{y} }{dt}=\omega_{0} j_{x} -2\lambda\sqrt{2\omega} x j_{z},
\end{equation*}
\begin{equation*}\label{eq: nJz2}
    \frac{d j_{z} }{dt}=2\lambda \sqrt{2\omega}  x j_{y},
\end{equation*}
\begin{equation*}\label{eq: nx_eq}
    \frac{d x }{dt}=  p-\frac{\kappa}{2} x, 
\end{equation*}
and
\begin{equation}\label{eq: np_eq}
    \frac{d p }{dt}=- \omega^{2}  x-\frac{\kappa}{2} p -2\lambda\sqrt{2\omega} j_{x}+16\alpha  \omega ^{2} x^{3}.
\end{equation}
In the above set of semiclassical dynamical equations we have employed a change of notation where we denote the expectation value of operator $A$, i.e., $\braket{A}$, by simply $A$.

We numerically solve these dynamical equations with the RK4 method, starting from the stable point defined by 
\begin{equation}
\begin{split}
x &=x_{0}, \\ 
p &=\frac{\kappa}{2}, \\ 
j_{x} & =\frac{16\alpha  \omega ^{2}x_{0}^{3}-\omega^{2}x_{0}-\frac{k^{2}}{4}x_{0}}{2\lambda\sqrt{2\omega}}, \\ 
j_{y} & =0,  \\ 
\text{and} \hspace{2em} 
j_{z} & =\omega_{0}\frac{16\alpha  \omega ^{2}x_{0}^{3}-\omega^{2}x_{0}-\frac{k^{2}}{4}x_{0}}{8\lambda^{2}\omega x_{0}}, \\ 
\text{with} \hspace{2em}
x_{0} & =-\frac{\sqrt{2\omega(1-(\frac{\omega_{0}(\omega^{2}+\frac{\kappa^{2}}{4})}{4\lambda^{2} \omega})^{2})}}{\omega^{2}+\frac{\kappa^{2}}{4}}.
\end{split} 
\label{eq:semiclassical-initial}
\end{equation}


We have solved the semiclassical dynamical equations, as described by Eq.~\eqref{eq: np_eq}, starting from the aforementioned fixed point, up to 1000 time-periods, for \(\kappa=0.05\). 
We have found out that for small detuning and weak anharmonicity, the semiclassical study shows time-translation symmetry breaking with period-doubling behaviour, as can be observed in Fig.~\ref{fig:semiclassical_plots}. 
We have also found this period-doubling behaviour to be robust over a range of detuning (\(\epsilon\)) and anharmonicity (\(\alpha\)), with the late-time oscillations sustained in phase, thereby signifying a DTC phase of the considered system.
For $\epsilon = 0, 0.01$ we get such a DTC phase in the range of $\alpha= 0-0.0025$. 
Since we assume the anharmonicity in the field mode to stem from transmons, we do not delve into the regime of negative $\alpha$ in this work. 
We have also uncovered that with higher detuning present in the system, the $\alpha$-range sustaining a 2DTC shrinks.

\subsection*{Extensivity of the Hamiltonian}
The Dicke model can be described through the Hamiltonian 
\begin{equation}\label{basic_ham}
    H=a^{\dag} a + \sum _{i} j^{z}_{i} + 2 \lambda \Big(\frac{a^{\dag}}{\sqrt{N}} + \frac{a}{\sqrt{N}} \Big)\sum _{i} j^x_i,
\end{equation}
where $j^{\mu}_i$ denotes pauli spin matrices of the $i^{th}$ qubit.
Eq.~\ref{basic_ham} can be rewritten in terms of the Hamiltonian density $h_i$ as
\begin{equation}\label{basic_ham_1}
    H=\sum_{i} h_{i},
\end{equation}
where 
\begin{equation}
    h_{i}=j^{z}_{i} + \frac{a^{\dag}}{\sqrt{N}} \frac{a}{\sqrt{N}} + 2 \lambda \Big(\frac{a^{\dag}}{\sqrt{N}} + \frac{a}{\sqrt{N}} \Big) j^x_i.
\end{equation}
We expect the energy to be an extensive quantity, i.e. it should scale linearly with the size of the system. 
On the other hand, $h_{i}$ should represent an intensive quantity. It is intuitive that the intensive quantities will be functions of $\frac{a}{\sqrt{N}}$ and $\frac{a^{\dagger}}{\sqrt{N}}$.
We write the Hamiltonian in terms of the mentioned density function, in the retrospective of the knowledge that $\frac{a}{\sqrt{N}}$ and $\frac{a^{\dagger}}{\sqrt{N}}$ turns out to be the meaningful quantities in the thermodynamic limit. 
As $\frac{a}{\sqrt{N}}$ and $\frac{a^{\dagger}}{\sqrt{N}}$ does not scale with N, we expect $h_i$ to be independent of $N$, i.e., the energy density to be intensive, while the Hamiltonian remains extensive.
Now, suppose we want to add a term like $(a^{\dag}+a)^{4}$ to the Dicke Hamiltonian. 
In that case, we should add $(\frac{a^{\dag}}{\sqrt{N}} +\frac{a}{\sqrt{N}})^{4}$ to $h_{i}$, to preserve the extensivity of the Hamiltonian, and thus rendering a Hamiltonian meaningful in the thermodynamic limit. 
Hence, the modified Hamiltonian should be
\begin{equation}
\begin{split}
H & =\sum _{i} \Big[j^{z}_{i} + \frac{a^{\dag}}{\sqrt{N}} \frac{a}{\sqrt{N}}+\Big(\frac{a^{\dag}}{\sqrt{N}}  +\frac{a}{\sqrt{N}}\Big)^{4} 
 \\ 
 & \hspace{10em} + 2 \lambda \Big(\frac{a^{\dag}}{\sqrt{N}} + \frac{a}{\sqrt{N}} \Big) j^x_i\Big] \\
&  = \sum _{i} j^{z}_{i} 
 + a^{\dag} a +\frac{1}{N}(a^{\dag}+a)^{4}\\
 & \hspace{8em} + 2 \lambda \Big(\frac{a^{\dag}}{\sqrt{N}} + \frac{a}{\sqrt{N}} \Big) \sum_i j^x_i,   
\end{split}
\end{equation}
which is the same as  the semiclassical Hamiltonian we have worked with, expressed via Eq.~\eqref{eq:semiclassical_hamiltonian}.

\begin{figure*}[t] 
\begin{minipage}{0.33\textwidth}
 \centering 
     \includegraphics[width=\textwidth]{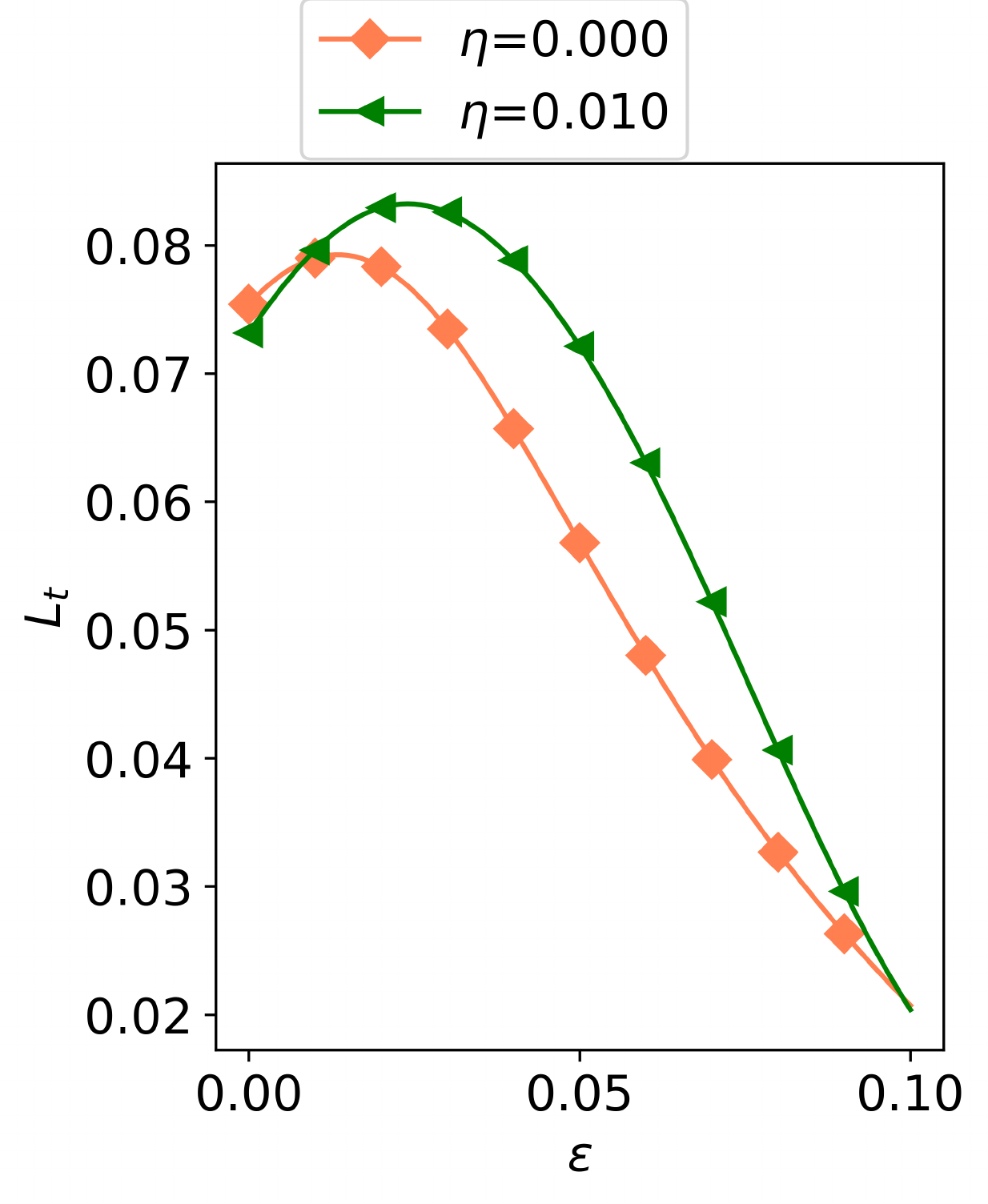}
    \centering
    \phantom{xxxxx} \textbf{(a)}
\end{minipage}
\hfill
\begin{minipage}{0.33\textwidth}
 \includegraphics[width=\textwidth]{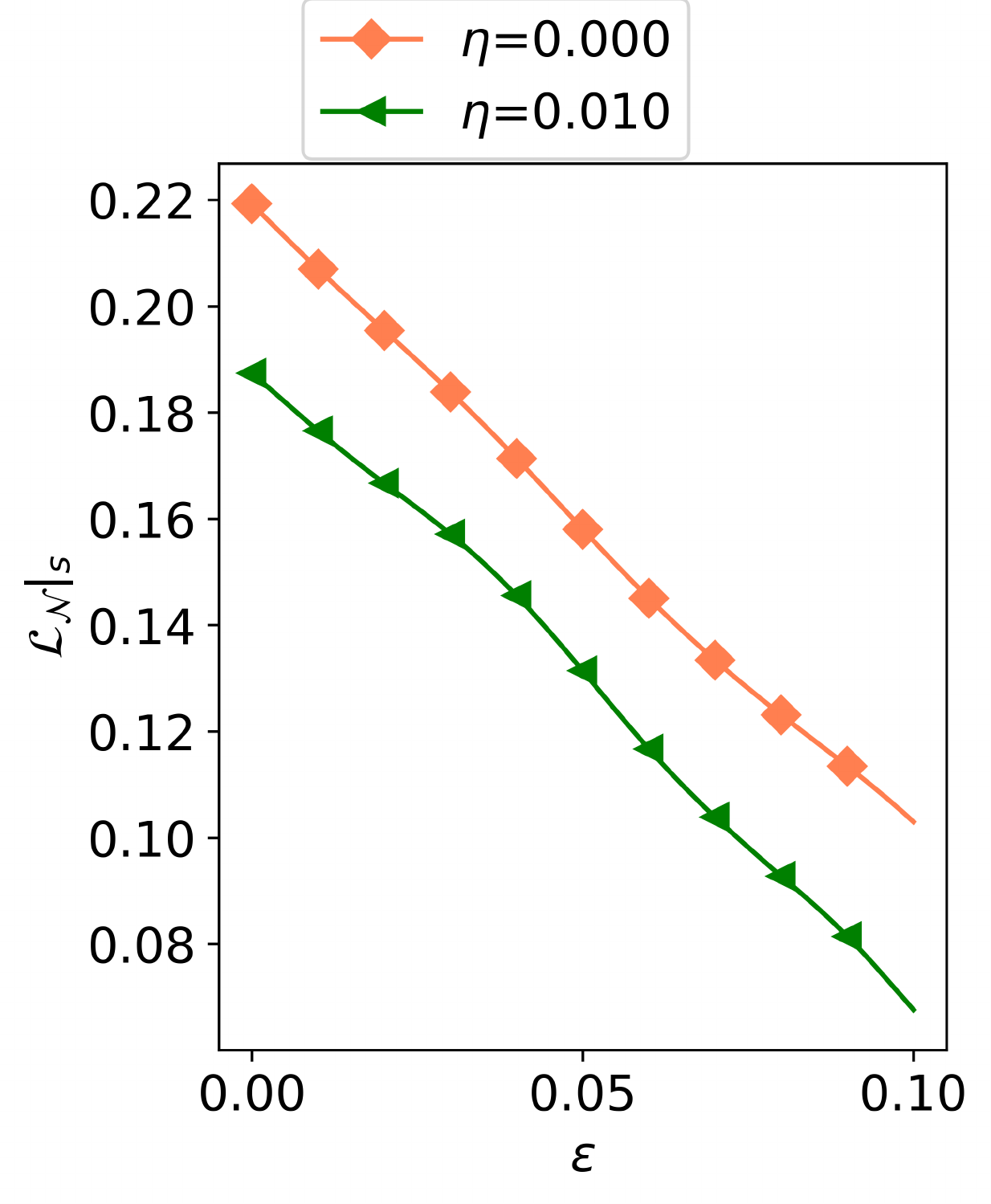}
    \centering
    \phantom{xxxxx} \textbf{(b)}
\end{minipage}
\begin{minipage}{0.33\textwidth}
 \includegraphics[width=\textwidth]{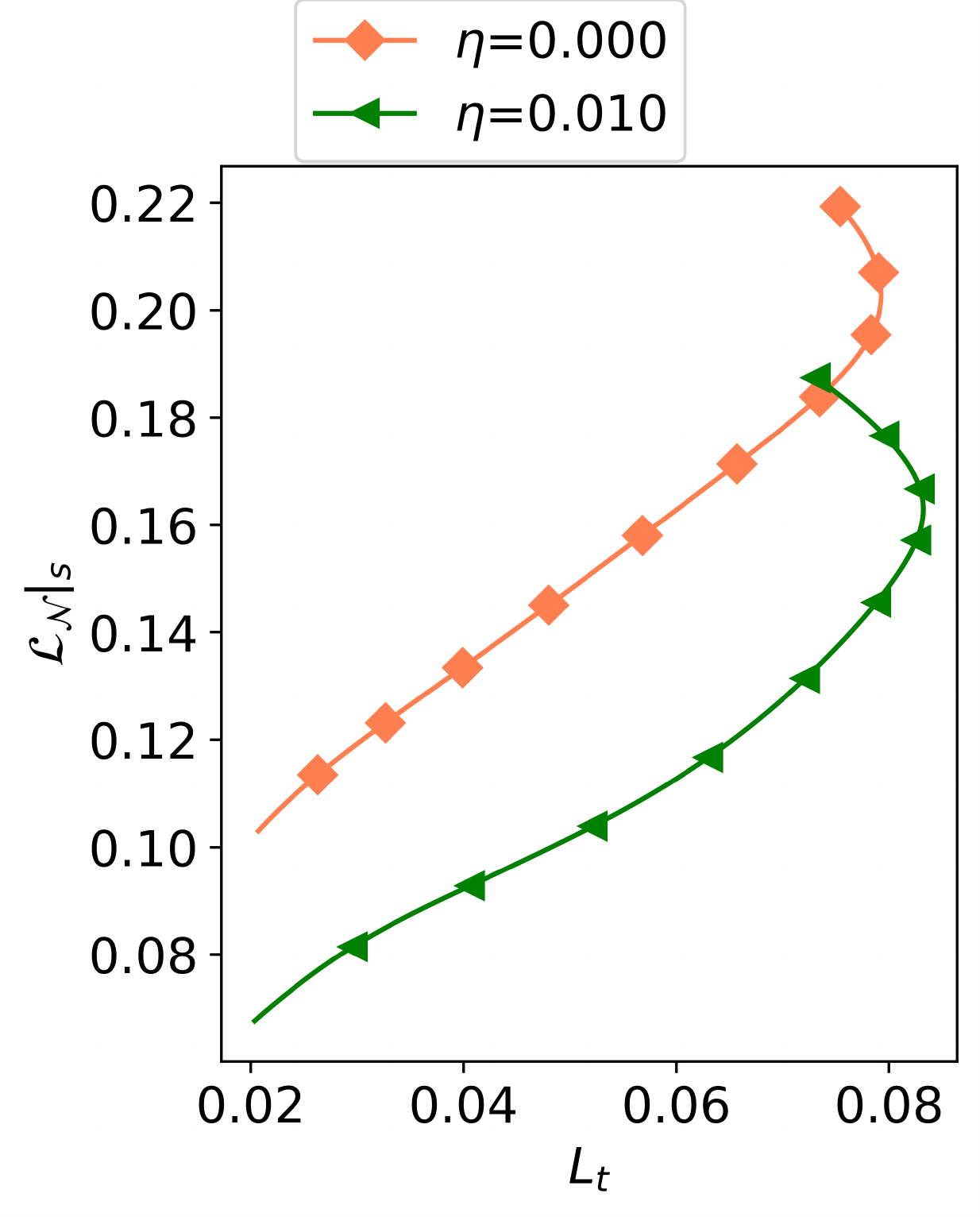}
    \centering
    \phantom{xxxxx} \textbf{(c)}
\end{minipage}
\caption{Dynamics of entanglement and DTC lifetime in a system of two qubits, driven by a photon/transmon field susceptible to dissipative leakage. Only the lowest 16 energy levels in the field mode are taken into account. In panels (a) and (b), detuning between the qubits and field mode, $\epsilon$ is plotted along the x-axis. Plotted along the y-axis of the panel (a) and the x-axis of the panel (c), is the lifetime of the DTC (denoted by $L_t$), as quantified via Eq.~\eqref{eq:lifetime}, scaled by $|(j_x)_0| \Delta_T/T$. We have set the other parameters according to $T_i=20T$, $\Delta_T = 80T$, $T = 2\pi/\omega_T$ , with $\omega_T=1$, in quantifier of DTC lifetime. $\kappa$ is chosen to be 0.05, enabling photon/transmon loss. Along the y-axes of panels (b) and (c), the long-time steady-state value of logarithmic negativity between the qubit-ensemble and field mode is plotted, in the unit of ebits, denoted by $\mathcal{L_N}|_s$. The above plots are simulated starting from the initial state $\ket{\Rightarrow}\otimes \ket{0}$, where $\ket{0}$ is the field vacuum and $\ket{\Rightarrow}$ is the eigenstate of $J_x$ operator with eigenvalue $N/2$. Plots in orange colour and with diamond markers describe the system driven by a lossy photon field. Green triangles feature the case when the driving field is a transmon mode, with $\eta=-0.01$. Note that $\eta$ is negative for transmon. The legends with the plots show the absolute values of the parameters.}
\label{fig:2q-lt-ent}
\rule{\hsize}{.1pt}
\end{figure*}

\begin{figure*}[t]
\begin{minipage}{0.33\textwidth}
 \centering 
     \includegraphics[width=\textwidth]{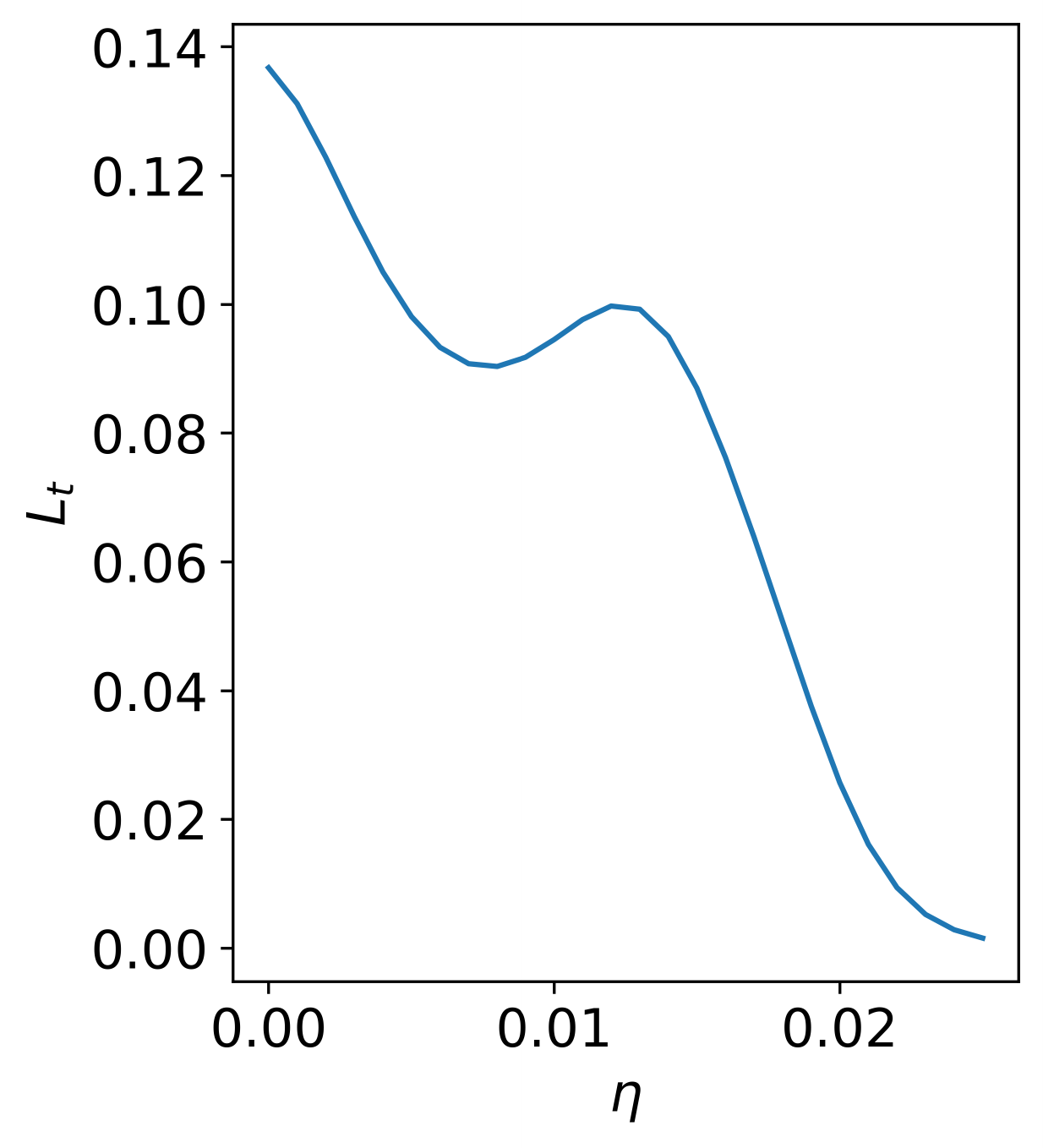}
    \centering
    \phantom{xxxxx} \textbf{(a)}
\end{minipage}
\hfill
\begin{minipage}{0.33\textwidth}
 \includegraphics[width=\textwidth]{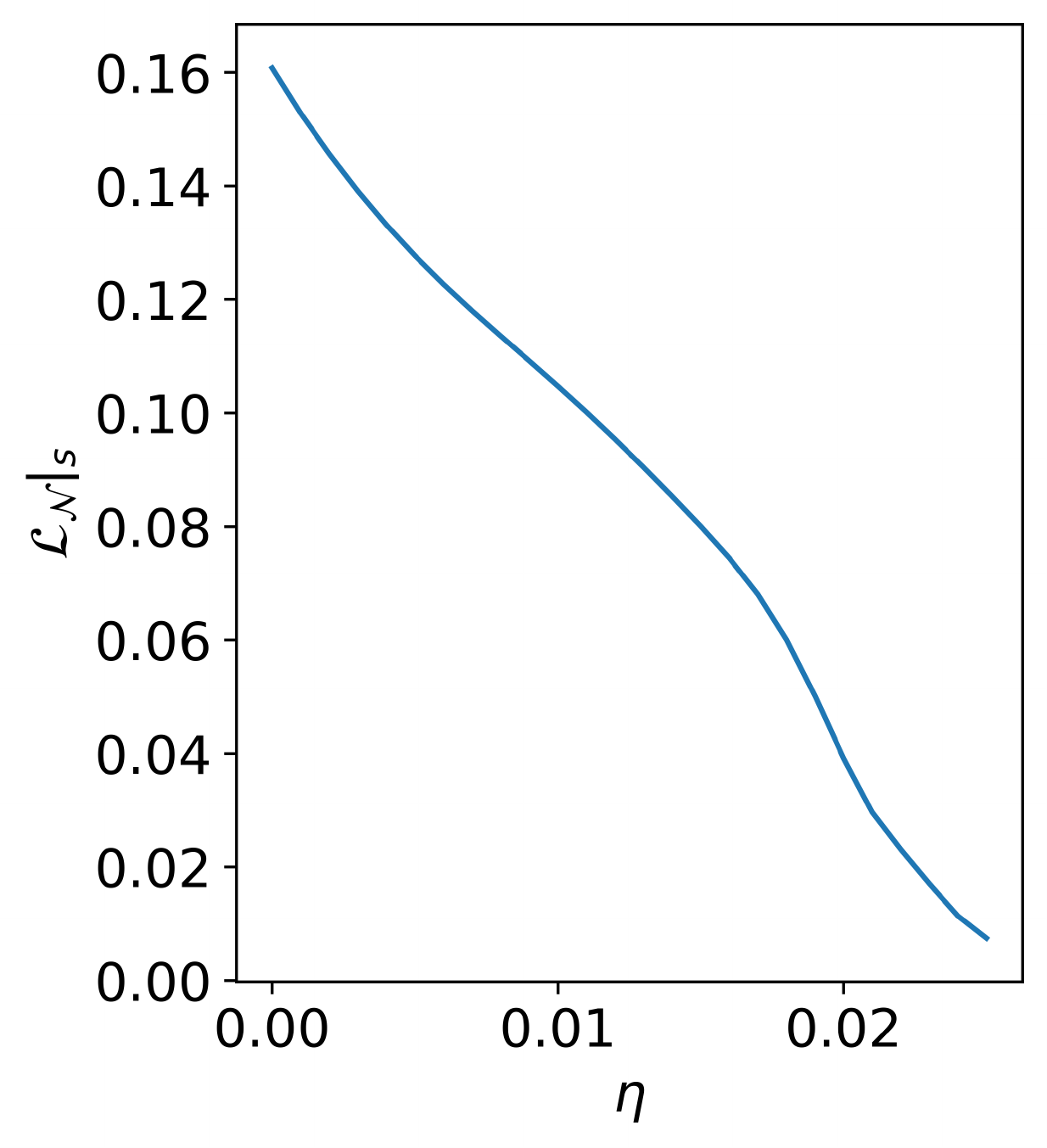}
    \centering
    \phantom{xxxxx} \textbf{(b)}
\end{minipage}
\begin{minipage}{0.33\textwidth}
 \includegraphics[width=\textwidth]{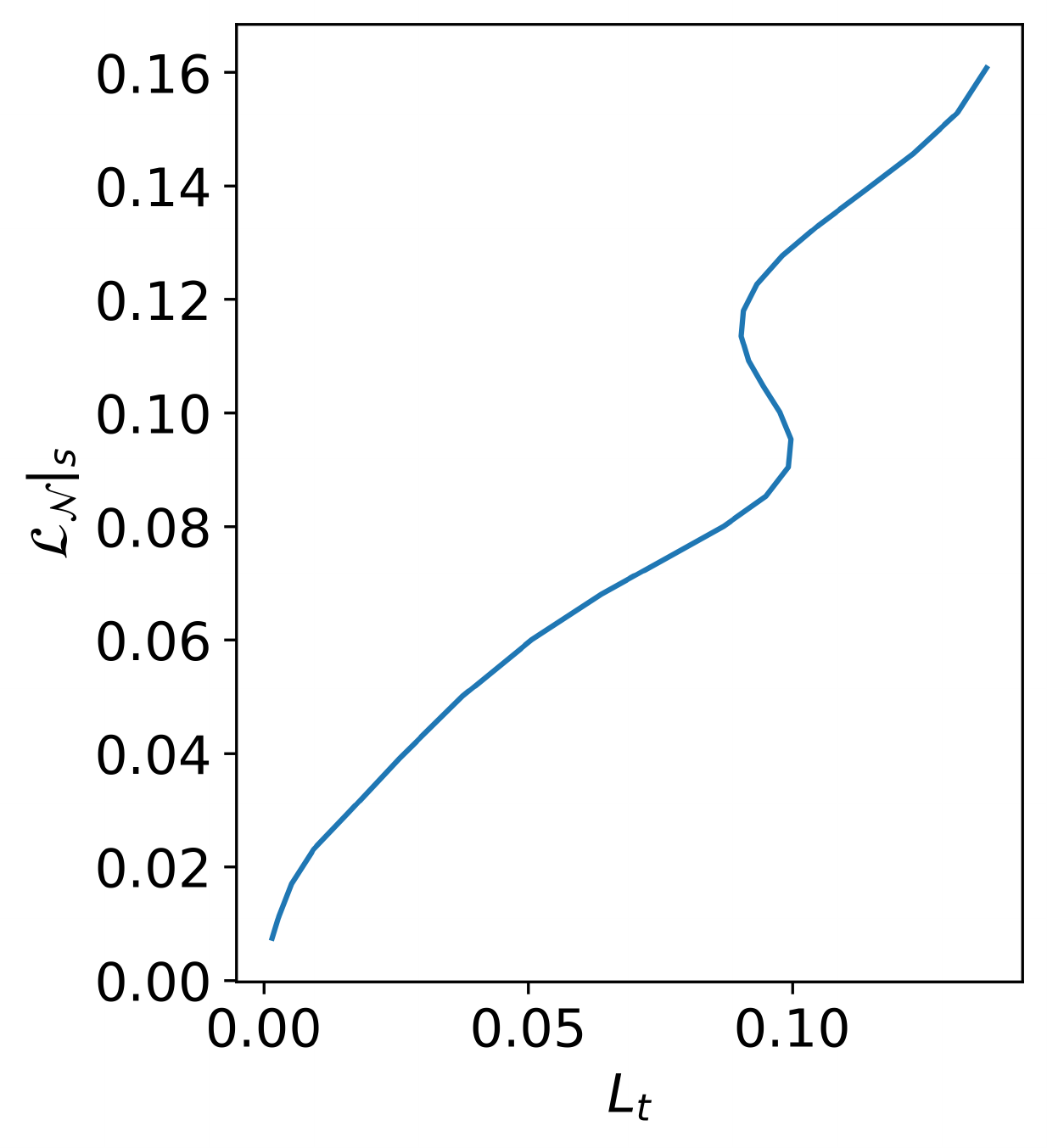}
    \centering
    \phantom{xxxxx} \textbf{(c)}
\end{minipage}
\caption{Dynamics of entanglement and DTC lifetime in a system of four qubits, driven by a transmon field susceptible to dissipative leakage. Only the lowest 16 energy levels in the field mode are taken into account. The anharmonicity $\eta$ is chosen to be negative and in the transmon regime.  x-axes of panels (a) and (b), exhibit this anharmonicity, but only in absolute values. Plotted along the y-axis of the panel (a) and the x-axis of the panel (c), is the lifetime of the DTC (denoted by $L_t$), as quantified via Eq.~\eqref{eq:lifetime}, scaled by $|(j_x)_0| \Delta_T/T$. We have set the parameters according to $T_i=20T$, $\Delta_T = 80T$, $T = 2\pi/\omega_T$ , with $\omega_T=1$, in quantifier of DTC lifetime. $\kappa$ is chosen to be 0.05, in effect of the dissipation from the field mode. Along the y-axes of panels (b) and (c), the long-time value of logarithmic negativity between the qubit-ensemble and field mode is plotted, in the unit of ebits, denoted by $\mathcal{L_N}|_s$. The above plots are simulated starting from the initial state $\ket{\Rightarrow}\otimes \ket{0}$, where $\ket{0}$ is the field vacuum and $\ket{\Rightarrow}$ is the eigenstate of $J_x$ operator with eigenvalue $N/2$. }
\label{fig:x-lt-ent-tmon-4q}
\rule{\hsize}{.1pt}
\end{figure*}

\section{Entanglement as  marker of lifetime of DTC} 
\label{sec:ent-lifetime}

In this section, we confine our investigation to the deep quantum regime of a few qubits to simulate the behaviour of the system in the presence of detuning and anharmonicity, as specified via Eq.~\eqref{eq:ME} and Eq.~\eqref{eq:H_s}. 
Throughout this work we stick to the strong coupling regime of $\lambda=1$, enabling the spontaneous symmetry breaking as described in the preceding section. 
The decay rate, due to the presence of a leaky cavity, is set to $ \kappa= 0.05 $. 
Within this parameter regime, the system can evade thermalisation despite being subjected to time-dependent drive and dissipation for a very long time, thereby ensuring a robust DTC phase with period-doubling dynamics.
For all the simulations illustrated in this section, we initialize the system in the state $\ket{\Rightarrow}\otimes \ket{0}$, where $\ket{0}$ is the photon/transmon vacuum and $\ket{\Rightarrow}$ is the eigenstate of $J_x$ operator with eigenvalue $N/2$. 

In this section we present our findings for a system made of two or four qubits and the field mode is truncated at 16 levels. 
Results for the three-qubit case are given in the appendix. 
We employ the $RK4$ method to solve the master equation in Eq.~\eqref{eq:ME} to capture the transient time evolution of this system, with Hamiltonian described by Eq.~\eqref{eq:H_s}. 
It is intriguing to observe that even with only two qubits, such a small system can capture the novelty of a DTC phase.
To explore this system further, we rove to investigate the entanglement in this system, and its possible connection to the stability and lifetime of such a novel non-equilibrium phase of matter. 
Simulations for the systems comprising two, three and four qubits reveal qualitatively similar features as described in the following. We only present a few representatives among them in this section.
We have used the measure of logarithmic negativity to capture the time evolution of the bipartite entanglement between the collection of two-level systems and the field mode.

\textbf{Logarithmic negativity:} One of the most widely used measures of bipartite entanglement is logarithmic negativity. Suppose, $\rho$ represents a bipartite density matrix comprising subsystems $A$ and $B$. 
Let us denote the partial transpose of $\rho$ on the subsystem $A$ by $\rho^{T_A}$. 
The negativity of the density matrix $\rho$ is defined by
\begin{equation}
    \mathcal{N}(\rho)=\frac{||\rho^{T_A}||_1-1}{2},
\end{equation}
where, $||A||_1=\text{Tr}\big(\sqrt{A^{\dagger}A}\big)$ stands for the trace norm of operator $A$. 
Here, $\mathcal{N}(\rho)$ is the absolute value of the sum of the negative eigenvalues of $\rho^{T_A}$. 
The logarithmic negativity of \(\rho\) is then defined as~\cite{Vidal2002, Plenio2005, Zyczkowski1998, Zyczkowski1999}
\begin{equation}
    \mathcal{L_N}(\rho) = \log_{2}(2 \mathcal{N}(\rho)+1).
\end{equation}
$\mathcal{L}_{\mathcal{N}}(\rho)$ equals 1 ebit for a maximally entangled state $\rho$. 
Here, an `ebit,' short for `entanglement bit,' is the unit of entanglement for bipartite quantum systems.

\subsection{Effect of detuning and anharmonicity on lifetime}
\label{subsec:effect-e,x}

Here we take up the pursuit to look into the stability of the time crystalline phase with varying detuning($\epsilon$) and anharmonicity($\eta$). We keep the anharmonicity in the field mode invariant to study the response of different important parameters, like the lifetime of the DTC phase and the entanglement in the system, to detuning. 
Similarly, we study the response of the system to change in anharmonicity, keeping the detuning fixed. 
In panels (a) and (b) of Fig.~\ref{fig:dtc-e} we can observe this response in the absence of anharmonicity, i.e., when the qubit chain is connected to the lossy cavity via a photon field. 
We can observe that the scaled angular momentum $j_x$  continues to show period-doubling behaviour, making this DTC phase robust against detuning.  
However, in Fig.~\ref{fig:dtc-e}(a), we can observe that with stronger detuning between the field and qubits, the system thermalises quicker, i.e., the lifetime of this transient DTC shortens. 
The corresponding bipartite entanglements (logarithmic negativity) in between the photon and qubit-chain are shown in Fig.~\ref{fig:dtc-e}(b). 
It is interesting to note that this entanglement stabilises to a constant finite positive value very early in the dynamics, and this stable value decreases with stronger detuning. 
Panels (c) and (d) of Fig.~\ref{fig:dtc-e} describe the same, except the field mode is a transmon (truncated at 16 levels) in this case. In this case, as well, we can clearly see that with increasing detuning the life-span of the DTC phase shortens and the entanglement between the field and qubit-chain saturates to a smaller value. 
It should be mentioned here that the value of $\eta=0.01$ keeps the anharmonic field well within the transmon regime.

Next, we study the response of the system when the detuning $\epsilon$ is absent, but we change the anharmonicity in the field mode from simple harmonic oscillators towards the transmon regime. 
Here also we can observe a robust 2DTC phase against changing anharmonicity in the system, as can be clearly concluded from Fig.~\ref{fig:dtc-x}(a). 
The same plot again demonstrated a shortened lifetime of this time crystalline phase with increasing anharmonicity $\eta$. 
Here as well, from Fig.~\ref{fig:dtc-x}(b), we can see that the entanglement between the field and qubit-chain quickly saturates to a steady positive finite value, the value of which decreases with an increase in $\eta$.

Fig.~\ref{fig:dtc-4} demonstrates similar findings for a system of four qubits driven by a photon/transmon field, prone to dissipative leakage.
We can observe that such a system shows transient 2DTC under the suitably chosen Floquet driving and dissipation, the lifetime of which shortens with increased anharmonicity in the field mode, in the absence of detuning, as described through Fig.~\ref{fig:dtc-4}(a). 
Fig.~\ref{fig:dtc-4}(b) shows the corresponding behaviours of the bipartite entanglement between the four qubits and the field mode. 
We can clearly observe the entanglement to saturate to a steady value very early in the dynamics, irrespective of the anharmonicity. 
We have checked that this entanglement value remains saturated for up to 500 driving periods, but data up to only 150 time periods is shown in the figure. 
From the same figure we can conclude that the mentioned saturated entanglement value decreases as we venture deeper into the anharmonic regime, i.e., larger $|\eta|$. 
In panels (c) and (d) of Fig.~\ref{fig:dtc-4}, we have illustrated the system's response to varied detuning, but in the presence of a photonic field, subjected to photon loss. 
Fig.~\ref{fig:dtc-4}(c) shows that the transient DTC thermalises early, as we subject the system to larger detuning. 
Here as well, we can observe that the same saturates the system to a smaller entanglement. 
All the figures described throughout this section clearly illustrate that both higher detuning and larger anharmonicity shorten the lifetime of the transient DTC, while the same results in the system saturating to a smaller entanglement between the qubits and the field.

\subsection{Emergence of entanglement as marker of lifetime}
\label{subsec:lt-ent}

In the preceding discussion, we have demonstrated how some changes in the system parameters can tweak the lifetime of the DTC phase of the system described. 
The entanglement between the field mode and the qubit-chain, within the studied parameter ranges, can also be indicative of the mentioned change, as revealed through the previous discussion. 
This interesting entanglement behaviour potentially predicts the DTC lifetime very early in the dynamics. 
Within the parameter space where the field mode is transmon and detuning is one order of magnitude smaller than the qubit energies, or less, we can perceive a very strong positive correlation between the lifetime of the DTC  and the aforementioned saturated value of entanglement. 
Motivated by this observation, we study this system further by expanding the parameter ranges to discover the validity and limitations of this conjecture, as depicted in the succeeding discussions. 

It should be mentioned here that we have repeated all the preceding investigations for systems with larger sizes, and we have found those results to be qualitatively the same as the ones demonstrated before.

\subsubsection*{Lifetime of a DTC phase} 

To investigate the correlation between the lifetime of a DTC and its entanglement dynamics, we have studied DTC in finite-sized driven-dissipative systems. 
As a result, the system is bound to thermalise beyond a finite time scale. 
The lifetime of this transient DTC can be quantified by how quickly the system thermalises.  
Clearly, the longer a system takes to thermalise, i.e., the longer the $j_x$($\braket{J_x/N}$) oscillations persist, the longer its lifetime is extended. 
So, to quantify the lifetime of a DTC phase, we sum over the amplitude of the late-time oscillations of $j_x$, measured stroboscopically, over a substantial range of time, multiplied with the driving period. 
This is basically twice the area under the graph showing stroboscopic oscillations in the $x-$component of magnetization, plotted stroboscopically, within the mentioned time-span.
The resultant quantity in normalised by the initial $j_x$-magnitude of the system and the number of driving periods contributing in the quantifier of DTC lifetime. 
Consequently, the lifetime ($L_t$) is quantified as

\begin{equation}
    L_t =  \frac{1}{\Delta_T}\sum\limits_{t=T_i}^{T_i+\Delta_T}\frac{|(j_x)_t|}{|(j_x)_0|} T.
    \label{eq:lifetime}
\end{equation}
It must be mentioned here that this $T_i$ in Eq.~\eqref{eq:lifetime} should preferably be chosen beyond $5T$, since the system takes $t \sim 5T $ time to start oscillating with a period of $2T$. 
We have typically chosen $T_i= 20T$ and $\Delta_T= 80T$, for most of the results presented in this work. 

We have also verified that if we quantify the lifetime of this transient DTC via some threshold magnitude of $j_x$ oscillations, all the results presented in this paper remain qualitatively the same.

\subsubsection{Qubits connected to lossy cavity via photon field}

Here we investigate the modulated open Dicke model, without any anharmonicity in the field. 
Previous studies on similar systems have revealed the range of detuning, within which the DTC phase of the mentioned system is robust~\cite{dtc-gong-2018}. 
Motivated by the dynamical phase diagram obtained in such studies, we stick to the parameter regime of $\epsilon\in [0,0.1]$ to investigate this system in the deep quantum regime.

In Fig.~\ref{fig:2q-lt-ent}, we present our findings for a system of two qubits, driven by a photonic field, which in turn is subjected to dissipative leakage. 
The corresponding plots are in orange colour, and are marked by diamond markers
From Fig.~\ref{fig:2q-lt-ent}(a) we can see that the lifetime of the transient DTC decreases rapidly within stronger detuning, except for a very narrow range near resonance. 
The corresponding plot in Fig.~\ref{fig:2q-lt-ent}(b) discloses that the saturated entanglement between the photon and qubit-array decreases linearly as the detuning increases. 
These features together ensure this entanglement as a marker for the lifetime of this transient DTC phase for a wide range of detuning, as can be verified from Fig.~\ref{fig:2q-lt-ent}(c). 
The narrow region of non-linearity in this plot is due to the slightly enhanced lifetime near resonance.

\subsubsection{Qubits connected to lossy cavity via transmon field}
\label{subsec:tmon-dtc}

We now proceed to the case where two or more qubits are driven by a transmon field, prone to dissipative leakage, as described in section~\ref{sec:model}. 
The case with only two qubits is illustrated in Fig.~\ref{fig:2q-lt-ent}, in green lines with triangle markers. 
Here again, we observe that the saturated entanglement value is linearly decreasing with increasing detuning, as depicted via the panel (b) of Fig.~\ref{fig:2q-lt-ent}. 
Panel (a) of the same figure shows the lifetime of this DTC to be diminishing with stronger detuning, apart from a short regime. 
Consequently, we can see the described steady-entanglement to be a good descriptor of the lifetime of this transient DTC  over a wide range of parameters. 
The strong positive correlation between the two is elucidated via Fig.~\ref{fig:2q-lt-ent}(c). 
Here also we see a limited region with non-linear features due to a slightly enhanced lifetime in the vicinity of resonant condition, just like in the case with the photonic field.

The findings for a system of four qubits, driven by a lossy transmon field, are presented in Fig.~\ref{fig:x-lt-ent-tmon-4q}.
The parameters are chosen carefully to keep the field within the transmon regime but with zero detuning. 
Fig.~\ref{fig:x-lt-ent-tmon-4q}(a) shows the lifetime of the DTC to be steadily declining in response to stronger anharmonicity, except non-linear behaviour over a small regime. 
In Fig.~\ref{fig:x-lt-ent-tmon-4q}(b), we observe the saturated entanglement between the qubits and the field to linearly fall off in response to the same. 
As a result, these two quantities show a strong correspondence, as depicted through Fig.~\ref{fig:x-lt-ent-tmon-4q}(c). 
In this figure, we can observe that an enhanced lifetime of the DTC corresponds to a heightened value of steady entanglement value, except for a narrow regime. 
We found out the system to behave almost identically in the presence of small detuning, in response to changing anharmonicity in the field mode. 




\section{Conclusion}
\label{sec:conclusion}

In this work, we have studied an ensemble of qubits featuring a DTC, in response to Floquet driving via its interaction with a dissipative photon or transmon field. 
It is fascinating to observe that the studied model shows a signature of a DTC phase even with only two qubits. 
We have investigated the system both in the semiclassical limit and in the deep quantum regime of a few qubits, and 
found
the system to exhibit a robust DTC phase with period-doubling dynamics, in the regime of weak anharmonicity and small detuning.
Interestingly, the entanglement between the qubits and the photon or the transmon mode in such systems stabilises to a steady value very early in the dynamics and remains invariant thereafter. 
Moreover, we uncover that this long-time value of  entanglement can be an indicator of the lifespan of such transient DTC that the system has embarked upon, for a broad regime of parameters.

We believe that our study can throw light upon whether there exists a 
role of quantum correlations in achieving and sustaining novel out-of-equilibrium phases of matter.
Moreover, we are inclined to take such entanglement behaviour, maintaining a finite saturated value from very early on in the dynamics, to signify the achieving of crypto-equilibrium, a hallmark of a DTC. 
Further analysis,
especially
in the case of `prethermal' time crystals, may uncover more insights about the role that quantum correlations may play in a time crystal, where the time crystalline phase is bound to thermalise beyond a finite time-scale even in the thermodynamic limit.
Throughout this study we have kept the anharmonicity in the field mode weak enough so that the introduced non-linearity shifts the fixed points obtained in the open Dicke model only slightly, giving us 
the expected symmetry breaking and superradiance in the considered parameter regime. 
This presumption has been reinforced by the numerical studies in this paper, where we come across clear signatures of symmetry breaking, and consequently the attainment of a robust period-doubled DTC aka 2DTC.

\begin{acknowledgments}

We acknowledge computations performed using Armadillo~\cite{Sanderson2016, Sanderson1} and QIClib~\cite{QIClib}.  
We acknowledge partial support from the Department of Science and Technology, Government of India through the QuEST grant (grant number DST/ICPS/QUST/Theme-3/2019/120). 

\end{acknowledgments}

\appendix

\section{Derivation of semiclassical equations}
\label{appen:semiclassical}

Here we provide the details of the derivation of the semiclassical dynamical equations.
In the system considered throughout this work, the qubit ensemble is driven via its interaction with a photon/transmon field, which is prone to dissipative leakage, the dissipation rate being $\kappa$. Then evolution of the density matrix ($\rho$) is given by
\[ \frac{d\rho}{dt}=-i[H,\rho]+\kappa ( a\rho a^{\dag}-\frac{1}{2}\{a^{\dag}a,\rho \}). \]
Equivalently, one can write the dynamical equation for the expectation value of the operator ($A$) in the Heisenberg picture as
\begin{equation}\label{Heisenberg_equation_of_motion}
    \frac{d\langle A\rangle}{dt}=i\langle[H,A]\rangle+\kappa(\langle a^{\dag}Aa\rangle-\frac{1}{2}\langle\{a^{\dag}a,A\}\rangle).
\end{equation}
Using the above equation we got the following dynamic equations
\begin{equation}\label{eq: Jx1}
    \frac{d\langle J_{x}\rangle}{dt}=-\omega_{0}\langle J_{y}\rangle
\end{equation}
\begin{equation}\label{eq: Jy1}
    \frac{d\langle J_{y}\rangle}{dt}=\omega_{0}\langle J_{x}\rangle-\frac{2\lambda}{\sqrt{N}}\langle (a^{\dag}+a) J_{z}\rangle
\end{equation}
\begin{equation}\label{eq: Jz1}
    \frac{d\langle J_{z}\rangle}{dt}=\frac{2\lambda}{\sqrt{N}}\langle (a^{\dag}+a) J_{y}\rangle
\end{equation}
\begin{equation}\label{eq: a_eq}
    \frac{d\langle a \rangle}{dt}=-(i\omega+\frac{\kappa}{2})\langle a \rangle-\frac{2i\lambda}{\sqrt{N}}\langle J_{x}\rangle+4i\frac{\alpha}{N} \langle (a^{\dag}+a)^{3} \rangle
\end{equation}
\begin{equation}\label{eq: adag_eq}
    \frac{d\langle a^{\dag} \rangle}{dt}=-(-i\omega+\frac{\kappa}{2})\langle a^{\dag} \rangle+\frac{2i\lambda}{\sqrt{N}}\langle J_{x}\rangle-4i\frac{\alpha}{N} \langle (a^{\dag}+a)^{3} \rangle
\end{equation}
From Eq.~\ref{eq: a_eq} and Eq.~\ref{eq: adag_eq} we get
\begin{equation}\label{eq: apadag_eq}
    \frac{d\langle a+a^{\dag} \rangle}{dt}=\omega i \langle a^{\dag}-a\rangle-\frac{\kappa}{2}\langle a+a^{\dag} \rangle 
\end{equation}
and 
\begin{equation}\label{eq: adagma_eq}
    \frac{d\langle a^{\dag}-a \rangle}{dt}=i \omega \langle a+a^{\dag}\rangle-\frac{\kappa}{2}\langle a^{\dag}-a \rangle +\frac{4i\lambda}{\sqrt{N}}\langle J_{x}\rangle-8i\frac{\alpha}{N} \langle (a^{\dag}+a)^{3} \rangle
\end{equation}
Here we have used 
\[[a,a^{\dag}]=1\] and \[[J_{i},J_{j}]=\epsilon_{ijk}J_{k}.\]

In terms of the average magnetisation $j_{\mu}={\braket{J_{\mu}}}/{N}$, and $x={\braket{(a+a^{\dag})}}/{\sqrt{2N\omega}}$, $p=i{\braket{(a^{\dag}-a)}}/{\sqrt{2N/\omega}}$, the equations governing the semiclassical dynamics of the system becomes 
\begin{equation*}\label{eq: Jx2}
    \frac{d\langle j_{x}\rangle}{dt}=-\omega_{0}\langle j_{y}\rangle,
\end{equation*}
\begin{equation*}\label{eq: Jy2}
    \frac{d\langle j_{y} \rangle}{dt}=\omega_{0}\langle j_{x} \rangle-2\lambda\sqrt{2\omega}\langle x j_{z}\rangle,
\end{equation*}
\begin{equation*}\label{eq: Jz2}
    \frac{d\langle j_{z} \rangle}{dt}=2\lambda \sqrt{2\omega} \langle x j_{y}\rangle,
\end{equation*}
\begin{equation*}\label{eq: x_eq}
    \frac{d\langle x \rangle}{dt}= \langle p\rangle-\frac{\kappa}{2}\langle x \rangle, 
\end{equation*}
and 
\begin{equation*}\label{eq: p_eq}
    \frac{d\langle p \rangle}{dt}=- \omega^{2} \langle x\rangle-\frac{\kappa}{2}\langle p \rangle -2\lambda\sqrt{2\omega}\langle j_{x}\rangle+16\alpha  \omega ^{2}\langle x^{3} \rangle.
\end{equation*}
If I have two operators $A$ and $B$ then we can write $\langle AB \rangle$ equals to $\langle A\rangle \langle B\rangle$ plus fluctuations. If the $\langle A\rangle \langle B\rangle$ becomes big compared to the fluctuations for large system size, then we can throw the fluctuations at the large system size limit. 
With this approximation, the semiclassical equations~\eqref{eq: np_eq} follows.

\bibliography{References}

\end{document}